\documentclass[12pt]{article}
\usepackage{amsmath,amssymb,subfigure}
\usepackage[dvips]{epsfig}
\newtheorem{lem}{Lemma}[section]

\newtheorem{rem}[lem]{Remark}

\newtheorem{prop}[lem]{Proposition}
\newtheorem{theo}[lem]{Theorem}
\newcommand{\proof}{{\bf Proof}:\quad}
\newcommand{\nc}{\newcommand}
\newcommand{\rnc}{\renewcommand}
\nc{\beqn}{\begin{eqnarray*}}
\nc{\eeqn}{\end{eqnarray*}}
\nc{\be}{\begin{equation}}
\nc{\ee}{\end{equation}}
\nc{\beqa}{\begin{eqnarray}}
\nc{\eeqa}{\end{eqnarray}}
\rnc{\a}{\alpha}
\rnc{\d}{\delta}
\nc{\ga}{\gamma}
\nc{\lb}{\lambda}
\nc{\f}{\phi}
\nc{\p}{\psi}
\nc{\e}{\eta}
\rnc{\c}{\chi}
\nc{\s}{\sigma}
\rnc{\t}{\theta}
\nc{\om}{\omega}
\rnc{\P}{\Psi}
\nc{\G}{\Gamma}

\nc{\ra}{\rightarrow}
\nc{\Ra}{\Rightarrow}
\nc{\LRa}{\Leftrightarrow}
\nc{\lra}{\leftrightarrow}
\nc{\ot}{\otimes}

\nc{\mat}[4]{\left(\begin{array}{cc}#1&#2\\#3&#4
\end{array}\right)}

\nc{\ca}{\mathcal{A}}
\nc{\cf}{\mathcal{F}}
\nc{\cs}{\mathcal{S}}
\nc{\cR}{\mathcal{R}}
\nc{\Gt}{\tilde{\Gamma}}

\nc{\dorz}{\omega_r(z)}

\newcommand{\Om}{\Omega}

\newcommand{\resi}{\operatornamewithlimits{Res}_{z=\infty}}

\newcommand{\vc}{{\vec{c}}}
\newcommand{\vu}{{\vec{u}}}
\nc{\VU}{{\vec{U}}}

\newcommand{\Rn}{{\rm I\!R}}

\begin{document}

\begin{center}
{\huge Bifurcation diagram of a}
\end{center}
\begin{center}
{\huge one-parameter family of }
\end{center}
\begin{center}
 {\huge   dispersive waves}
\end{center}
\begin{center}
{\large T. Grava}\footnote{E-mail: grava@math.umd.edu}
\end{center}
\begin{center}
Department of Mathematics,
University of Maryland,\\
College Park 20742-4015 MD,
USA\\
and\\
SISSA, via Beirut 2-4 34014 Trieste, Italy
\end{center}
 
\begin{abstract}
\noindent The Korteweg de Vries (KdV) 
equation with small dispersion is a model for the
formation and propagation of dispersive shock waves
in one dimension. Dispersive shock
waves in KdV  are characterized by the appearance of zones of 
rapid modulated oscillations in the solution of the Cauchy problem 
with smooth  initial data.
  The modulation in time and space of the amplitudes,
the frequencies and the wave-numbers of these oscillations 
and their interactions is approximately described
by the $g$-phase Whitham equations.
 We study
the initial value problem for the Whitham  equations for a one parameter
 family of monotone decreasing
initial data. We obtain the bifurcation diagram of the  
number $g$ of interacting oscillatory zones.
\end{abstract}

\section{ Introduction}
The solution of the Cauchy problem for the KdV  equation 
\begin{eqnarray}
\left\{\begin{array}{lll}
u_t+6uu_x+\epsilon^2u_{xxx}=0, \quad t,x\in\Rn &&\\
&&\\
u(x,t=0,\epsilon)=u_0(x)&&
\label{a1}
\end{array}\right.
\end{eqnarray}
where  $\epsilon>0$ is a small parameter, 
 is characterized by the appearance of  zones of
rapid modulated oscillations. These modulated 
oscillations are called dispersive
shock waves and they are described in terms of the Whitham equations.
For $\epsilon>0$, no matter how small, the solution of (\ref{a1}) 
with smooth  initial data exists and remains smooth for all $t>0$. 
For $\epsilon= 0$  (\ref{a1}) becomes the Cauchy problem for 
the Burgers equation $u_t+6uu_x=0$. If the initial data 
$u_0(x)$ is decreasing somewhere the solution $u(x,t)$ of 
the Burgers equation 
has  always  a point  $(x_0,t_0)$  of gradient catastrophe  
where  an infinite derivative develops.  

\begin{figure}[htb] 
\centerline{\epsfig{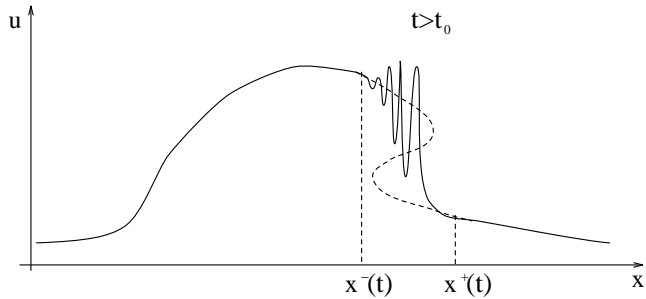}}
\caption{The dashed line represents the formal solution of the 
Burgers equation after the time of gradient catastrophe $t=t_0$. 
The oscillations on the picture are 
 close to a modulated periodic wave.\label{mulosc}}
\end{figure}

\noindent
After the time of gradient catastrophe
the solution
$u(x,t,\epsilon)$ of (\ref{a1}) develops an expanding region filled with
rapid modulated oscillations as shown in Fig.~\ref{mulosc}.

The idea and first example of the  description of the 
dispersive shock 
waves were
proposed by the physicists Gurevich and Pitaevski (GP) \cite{GP}. 
 They  considered the oscillatory zone to be approximately
 described by a modulated periodic wave:
\begin{equation}
\label{cnoidal}
u(x,t,\epsilon)\simeq \dfrac{2a}{s} dn^2\left[(\dfrac{a}{6s})^{1/2}
\dfrac{(x-Vt+x_0)}{\epsilon},\;s\,\right]+
\gamma\,,
\end{equation}
where $dn(y,s)$ is the Jacobi elliptic function of modulus 
$s\in(0,1)$, $x_0$ is a suitable phase,  the quantities
$\;a$, $\;s$,  $\;\gamma$ and
 $V=\left[2a\dfrac{2-s}{3s}+\gamma\right]$
depend   on  $x$ and $t$.
These quantities evolve according to the Whitham equations \cite{W}
which guarantee the validity of the approximate description
 (\ref{cnoidal}). For constant values of the  parameters 
$a$,$\;s$ and $\gamma$, $\;u(x,t,\epsilon)$ is an exact periodic 
solution of KdV with amplitude $a$,  
 wave number $k$ 
and  frequency  $\omega$  given by the relations
\begin{equation*}
 a=u_{max}(x,t,\epsilon)-u_{min}(x,t,\epsilon)\,,\quad\quad 
k = \dfrac{\pi}{\epsilon K(s)}\sqrt{\dfrac{a}{6s}}\,,\quad  \quad
\omega=\dfrac{k}{V}\,,
\end{equation*}
where $K(s)$ is the elliptic integral of the first 
kind of modulus $s$.

\noindent
 Whitham
introduced the Riemann invariants $u_1> u_2> u_3$ to 
write the  equations
for $a$, $\;s$ and $\gamma$ in diagonal form. These quantities  
 are expressed in terms
of  $u_1>u_2>u_3$   by the relations
\begin{equation}
a=u_2-u_3\,\quad\; s=\dfrac{u_2-u_3}{u_1-u_3}\,,\quad\; 
\gamma=u_2+u_3-u_1\,.
\end{equation}
The Whitham  equations for the $u_i$, $i=1,2,3$  read \cite{W}
\begin{equation}
\label{whitham}
\partial_t u_{i}(x,t)+\lambda_i(u_1,u_2,u_3) 
\partial_x u_{i}(x,t)=0,
\quad i=1,2,3\,,
\end{equation}
where
\begin{equation}
\label{l1}
\lambda_i(u_1,u_2,u_3)=2(u_1+u_2+u_3)+
\dfrac{\prod_{j\neq i,j=1}^3(u_i-u_j)}{u_i+\alpha_0},\quad 
\end{equation}
\begin{equation}
\nonumber
\alpha_0=-u_1+(u_1-u_3)\dfrac{E(s)}{K(s)}
\end{equation}
and $E(s)$ is the complete elliptic integral of the second kind.
\begin{figure}[tbh]
\centerline{\epsfig{figure=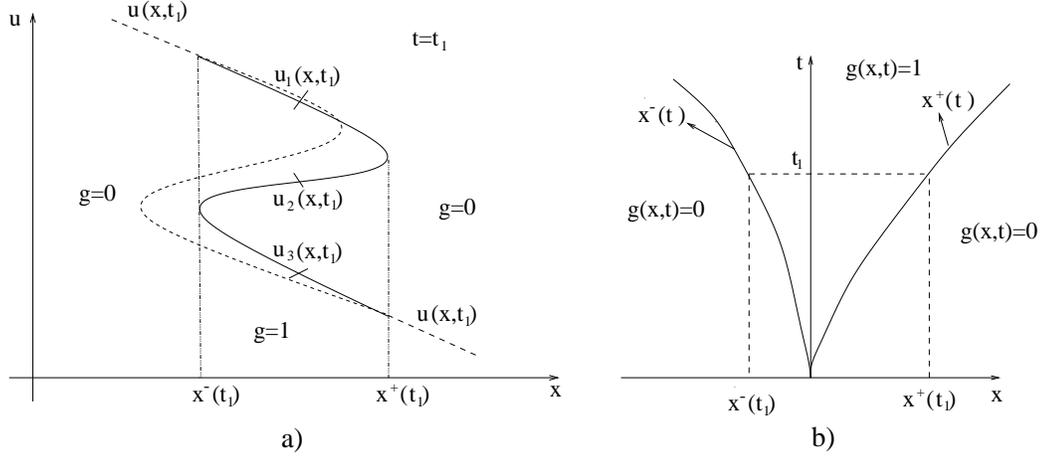,height=6cm}}
\caption{On   picture   a)  
the dashed line represents the formal solution
of the Burgers equation, the continuous line represents the
 solution of the Whitham equations. The solution
$(u_1(x,t)\,,\,u_2(x,t),u_3(x,t))$
of the Whitham equations  and the position of the boundaries 
$x^-(t)$, $x^+(t)$ of the 
 oscillatory  zone are to be determined from the condition
$u(x^-(t),t)=u_1(x^-(t),t),\,\,$  
$u_x(x^-(t),t)=u_{1x}(x^-(t),t),\,\,$
$u(x^+(t),t)=u_3(x^+(t),t),\,$ $u_x(x^+(t),t)=u_{3x}(x^+(t),t)$
where $u(x,t)$ is the solution of the Burgers equation.
 Picture b) represents the functions $x^-(t)$ and $x^+(t)$  
on the $x-t$ plane.
\label{ivp} }
\end{figure}

\noindent
The variable $u_2$ can vary  from $u_3$ to $u_1$. 
The oscillatory region is bounded  on one side by the point
$x^-(t)$  where $u_2(x,t)=u_3(x,t)$, and on the other side by the
point $x^+(t)$ where $u_2(x,t)=u_1(x,t)$ (see Fig.~\ref{mulosc}
 and Fig.~\ref{ivp}).  Outside the region $(x^-(t), x^+(t))$
the solution $u(x,t,\epsilon)$ of (\ref{a1}) is well approximated by
the solution $u(x,t)$  of the Burgers equation. 

\noindent
In the theory of singularity a function behaves
 quadratically in the neighborhood of its generic
 singular point \cite{Ar}. In a similar 
 way,  in the theory of dispersive shock
        waves
 the generic analytic monotone decreasing initial data 
$x= f\left(u|_{t=0}\right)$
         behaves (up to shifts and rescalings) like $x=-u^3$ in 
        the neighborhood of its  generic breaking  point. 
Thus the  behavior of the generic analytic initial data 
near a point of gradient catastrophe is  
\begin{equation}
\label{Taylor}
x-x_0\simeq- c_3(u-u_0)^3-c_4(u-u_0)^4-\dots-c_k(u-u_0)^k-\dots.
\end{equation}
GP kept only the first term of  the above Taylor series namely they
 considered the initial data $x-x_0=-(u-u_0)^3$ (up to rescaling)
 and solved
 the  equations (\ref{whitham})  numerically for such  initial data.
 Potemin \cite{P} obtained the analytic solution 
of the equations (\ref{whitham}) for the same initial data 
and  showed that 
 $x^-(t)=x_0-12\sqrt3\, (t-t_0)^{\frac{3}{2}}$ and
$x^+(t)=x_0+4/3\sqrt{5/3}\,(t-t_0)^{\frac{3}{2}}$ 
 (see Fig.~\ref{ivp}).

\noindent 
The higher order terms of the Taylor expansion (\ref{Taylor})
 are negligible only for $t\ll t_0+2\sqrt 3\sqrt{\frac{c_3}{c_4}}$ 
\cite{M}.  When one starts considering the higher order terms 
of the expansion (\ref{Taylor}), we come to a situation in
which two or more oscillatory wave trains come in interaction
or in the case in which the solution of (\ref{whitham}) comes
itself to a point of gradient catastrophe.
We will see that such a phenomena necessarily occur  
in the simplest  non  trivial deformation of the cubic law
which turns out to be 
 the fifth order  polynomial:
\beqa
\label{gd}
x\simeq -c_3(u-u_0)^3-c_4(u-u_0)^4-c_5(u-u_0)^5\,, 
\quad c_3>0,\;c_5> 0\;.
\eeqa
Through the shift $(x\rightarrow x+6tu_0,u\rightarrow u+u_0)$ 
the above initial data becomes 
$x\simeq -c_3 u^3-c_4 u^4-c_5 u^5\,.$
The  parameter space  $(c_3,c_4,c_5)$ can be reduced
exploiting the invariance of  the 
 KdV equation under the   groups of transformations 
$(x\rightarrow k^3\,x,\;t\rightarrow k^2\,t,\;
 u\rightarrow k\,u)$, $k\neq 0 $  and 
$(x\rightarrow  \alpha\,x, \; t\rightarrow \alpha \,t)$, 
$\alpha\neq 0$.
These transformations change, however, the value of the 
small parameter $\epsilon$.
Taking $k=\sqrt{c_3/c_5}$ and $\alpha=c_3$,
the initial data (\ref{gd}) can be reduced to the form
\beqa
\label{bi}
x=- u^3- c\, u^4-u^5\;,
\eeqa
where the dimensionless parameter $c$ is chosen in the form
\beqn
c=\dfrac{c_4}{\sqrt{c_3\,c_5}}=
\dfrac{\sqrt 5}{2}\dfrac{f^{IV}(\xi)}{\sqrt{
 f^{\prime\prime\prime}(\xi)f^{V}(\xi)}}\,.
\eeqn 
The monotonicity condition requires $c^2\le 15/4$.
From the above considerations the polynomial (\ref{bi})
represents the  generic one-parameter 
deformation of a monotone decreasing analytic  initial data
with cubic inflection point. 

\noindent
For cubic initial data  the solution $u(x,t,\epsilon)$ of 
(\ref{a1}) is approximately given by
equation (\ref{cnoidal}) inside the oscillatory zone  and by
 the solution of
the Burgers equation outside this zone. For general initial data 
one have to use higher genera analogues of the Whitham equations 
(\ref{whitham})
to approximately described the solution 
$u(x,t,\epsilon)$ of (\ref{a1})
\cite{LL},\cite{V}. 
Lax and Levermore \cite{LL}, and Venakides \cite{V}  described,
for certain particular classes of initial data,  the 
dispersive shock waves 
in the frame of the zero-dispersion asymptotics  for 
the solution of the inverse scattering problem  of KdV.
According to their results,   to the solution
$u(x,t,\epsilon)$ as $\epsilon\rightarrow 0$ it corresponds a 
decomposition of the  $(x,t)$
plane into a number of domains $D_g$, $g=0,1,\dots$. In the 
domain $D_g$  the principal term of the
asymptotics is given by the $g$-phase solution of the KdV 
equation \cite{DMN}
\begin{equation}
\label{asymp}
u(x,t,\epsilon)\cong\Phi\left(\frac{S_1(x,t)}{\epsilon},\dots,
\frac{S_g(x,t)}{\epsilon};\,u_1(x,t),\dots,
u_{2g+1}(x,t)\right)\,,
\end{equation}
where the functions $S_j(x,t)$  satisfy the equations \cite{DN}
\begin{equation}
\label{wavep}
\dfrac{\partial S_j}{\partial x}=k_j(\vu(x,t))\,,\quad
\dfrac{\partial S_j}{\partial t}=\omega_j(\vu(x,t))\,,\;\;
 j=1,\dots,g\,,
\end{equation}
and the formula
\[
u(x,t)= \Phi\left(k_1x+\omega_1 t+\phi_1,\dots,k_g x+
\omega_g t+\phi_g\,;\,u_1,\dots,u_{2g+1}\right)
\]
for constant values of  the parameters $u_1,\dots,u_{2g+1}$, 
$k_j=k_j(\vu)$ and $\omega_j=\omega_j(\vu)$ and for arbitrary 
$\phi_j$, $j=1,\dots,g$, 
gives the family of the so-called $g$-phase
 exact solutions of KdV for
$\epsilon=1$ \cite{DMN}. 
The wave parameters in (\ref{wavep}) depend on the functions
 $u_1(x,t)>\dots>u_{2g+1}(x,t)$
which satisfy the system of equations 
\beqa
\label{whithamg}
\dfrac{\partial u_i}{\partial t}+
\lambda_i(u_1,u_2,\dots,u_{2g+1})\dfrac{
\partial u_i}{\partial x}=0\;,\quad i=1,...,2g+1\;,\quad g\ge 0\;.
\label{a2}
\eeqa 
For a given $g$ the system (\ref{whithamg}) is called $g$-phase
 Whitham equations. The Whitham equations is the collection of all 
systems (\ref{whithamg}) for $g\geq 0$. The zero-phase Whitham
 equation coincides with the Burgers equation.
For $g>0$ the speeds $\lambda_i(u_1,u_2,\dots,u_{2g+1})$,
$i=1,2,\dots,2g+1$, depend on  $u_1,\dots,u_{2g+1},$ through  
complete hyperelliptic integrals  of genus $g$. For this reason the
$g$-phase solution is also called solution of genus $g$.
When the $g$-phase solution comes
to a point of gradient catastrophe, it should be continued
by the $(g+1)$-phase solution.
The main problem  is to glue together the
solutions of the Whitham equations for different
 $g$ in order to produce a $C^1$-smooth curve in the $x-u$ plane
 evolving smoothly with $t$ . This problem is referred to 
as initial value problem 
of the Whitham  equations. \noindent The genus $g(x,t)$ is 
a piecewise
constant  on the $(x,t)$ 
plane (see Fig.~\ref{ivp}).
 For generic initial data it is not not whether the genus 
$g(x,t)$ is
a bounded  function.

\noindent
The algebraic geometric description of these equations for $g>1$,  
was first derived in \cite{FFM}.
Dubrovin and Novikov  developed a geometric Hamiltonian theory for
the Whitham equations \cite{DN}.
 Based on this theory, Tsarev \cite{T} was able to prove
that, for each $g$, equations (\ref{a2}) 
 can be solved by a generalized  method of characteristic. 
This method was put into an algebro-geometric setting 
by Krichever \cite{K}. 
Recently Dubrovin  proposed a variational principle for the Whitham
equations. 
In \cite{D1} the minimizer of a functional defined  on a certain 
infinite-dimensional space formally solves the initial value 
problem  of the Whitham equations for each point of the 
$(x,t)$ plane.

\noindent
We use this variational principle to arrive at the main result of
this paper. This is a complete description of the bifurcation 
diagram  in the $x-t$ plane
 of  the    one-parameter family of initial value problems
for  (\ref{whithamg}).
 For the purpose we use the result in \cite{G} where it is shown that
for polynomial initial data of degree $2N+1$ the solution of the 
Whitham equations has genus at most equal to $N$.
In our  case of fifth degree polynomial initial data, 
 the  solution of the initial value problem turns 
out to be glued from solutions of
the Whitham equations of  genera $g=0$, $g=1$ and $g=2$.
 The various topological type of
bifurcation diagrams are described in terms of the 
number of points of
gradient catastrophe that appear in the solution of the Whitham
equation and other particular points which we call
double-leading edge, double trailing edge and leading-trailing
edge. 

\noindent
This paper is organized as follows. On Sec. 2 we give some basic
results  about the theory of the Whitham equations and we describe  
Dubrovin's variational principle for the Whitham equations. 
In Sec. 3 we describe the  phase transition boundaries 
 for general initial data. In Sec. 4 we study the function 
$g=g(x,t)$ for the one parameter family of initial data (\ref{bi}).


\section{Preliminaries   on the theory of the Whitham
 equations}
\setcounter{equation}{0}
\setcounter{figure}{0}
The $g$-phase Whitham equations (\ref{a2}) are constructed 
in the following way.

\noindent
On  the Riemann surface 
\beqa
\label{b5}
\cs_g:=\left\{(P,\mu(r)),\;\mu^2=
\prod_{i=1}^{2g+1}(r-u_i)\right\}\;,\;\;
u_1>u_2>\dots>u_{2g+1},\;\; u_i\in\Rn,
\eeqa
we define the  abelian differentials of the  second kind
 $\s_k$, $k\geq 0$ with a pole at infinity of  order  
$2k+2$ \cite{Sp}
\beqa
\label{ds}
\s^g_{k}(r)=\dfrac{P^g_{k}(r,\vu)}{2\mu}dr\;,\quad
P^g_{k}(r,\vu)=r^{k+g}+\alpha_{k+g-1} 
r^{k+g-1}+\dots+\alpha_0\,,
\eeqa
where $\vu=(u_1,\dots,u_{2g+1})$ and the constants 
$\alpha_i=\alpha_i(\vu)$, $i=0,1,\dots,k+g-1\,,$   
are uniquely  determined by the conditions:
\beqa
\label{n2}
\int_{u_{2j+1}}^{u_{2j}} \s^g_{k}(r)=0,\quad j=1,2,\dots,g\,,
\quad
\s^g_{k}(r)= \left[\dfrac{r^{k-\frac{1}{2}}}{2}
+O(r^{-\frac{3}{2}})\right]dr\;\;\mbox{for large }
\mid r\mid\,.
\eeqa
We use the notation 
\begin{equation}
\label{dp}
dp_g(r)=\s^g_0(r),\quad dq_g(r)=12\s^g_1(r)\,.
\end{equation}
In literature the differentials $dp_g(r)$ and $dq_g(r)$ are
 called quasi momentum and
quasi-energy respectively \cite{DN}.
The speeds $\lambda_i(\vu)$ of the $g$-phase Whitham 
equations (\ref{a2})  are given
by the ratio \cite{FFM}:
\beqa
\label{b9}
\lambda_i(\vu)=\left.\dfrac{dq_g(r)}{dp_g(r)}\right|_{r=u_i}\;,
\quad i=1,2\dots 2g+1\,.
\eeqa
In the case $g=0$,  we have   $dp_0(r)=\dfrac{dr}{\sqrt{r-u}}$ and
$dq_0(r)=\dfrac{12r-6u}{\sqrt{r-u}}dr\;$ respectively,
so that the zero-phase Whitham equation (\ref{a2}) coincides
 with the Burgers equation
$u_t+6uu_x=0\,.$
We consider  monotone decreasing polynomial initial data  of 
the form
\beqa
\label{b18}
x=f(u)=c_0+c_1 u+\dots+c_{2N}u^{2N}+c_{2N+1} u^{2N+1}\;.
\eeqa
For such initial data the solution of the Burgers equation  
is obtained by
the method of characteristic \cite{W} and is given by the
 expression
\beqa
\label{burgers}
x=6t\,u+f(u)\,,
\eeqa
This solution is globally well defined only for  $0\leq t <t_c$ 
where the time  
$t_c=\frac{1}{6}\min_{u\in\Rn}[-f^{\prime}(u)]$ is the time of
gradient catastrophe of the solution. The breaking is cause by an 
inflection point in the initial data. 
For $t>t_c$ the solution of the Whitham equations is 
 obtained gluing together
$C^1$-smoothly solutions of different genera. 
The $g$-phase solution $u_1>u_2>\dots >u_{2g+1}$ 
which is  attached $C^1$ smoothly to the $(g+1)$-phase  
solution  is  obtained by a generalization of the method of
 characteristic and reads
 \cite{T},\cite{FRT2}   
\begin{equation}
\label{ch}
x=\lambda_i(\vu)\,t+w_i(\vu),\quad 
w_i(\vu)=\left.\dfrac{ds_g(r)}{dp_g(r)}\right|_{r=u_i}\,,
\quad i=1,\dots ,2g+1\,,
\end{equation}
where $\lambda_i(\vu)$ has been defined in (\ref{b9}) and 
the differential $ds_g(r)$ is given by the expression
\begin{equation}
\label{dsg}
ds_g(r)=\sum_{k=0}^{2N+1} \dfrac{2^k k!}{(2k-1)!!} c_k \s^g_k(r)\,,
\end{equation}
where $\s^g_k$ has been defined in (\ref{ds}) and the $c_k$'s have
been defined in (\ref{b18}).

\noindent
The solution (\ref{ch})  can 
 be written in the equivalent  algebro-geometric
 form. Consider the differential
\begin{equation}
\label{omega}
\Om_g(r)=xdp_g(r)-tdq_g(r)-ds_g(r)\,,
\end{equation}
where $dp_g$, $dq_g$ and $ds_g$  have been defined 
in (\ref{dp}) and (\ref{dsg}) respectively. 
Then the $g$-phase solution is given by the relation \cite{K,FRT2}
\begin{equation}
\label{b20}
\Om_g(r)\mid_{r=u_i}=0\;,\quad i=1,2\dots,2g+1\,.
\end{equation}

\noindent
The solution $u_1>u_2>\dots >u_{2g+1}$  of the $g$-phase Whitham
equations (\ref{whithamg})
 is  implicitly defined
 as a function of $x$ and $t$ by the equations  
(\ref{ch}) or (\ref{b20}).
The solution is uniquely defined 
only for those  $x$ and $t$ such that the functions 
$u_i(x,t)$ are real and $\partial_xu_i(x,t)$, $i=1,\dots,2g+1$, 
 are not vanishing. 

\subsection{Variational principle for the Whitham equations}
The solution of  the Whitham equations, for a given initial data,
can be written as the minimizer of a functional defined
on a certain infinite-dimensional space \cite{D1}.
Let us first consider the zero-phase equation. 
The characteristic equation  $x=6tu+f(u)$, 
where $f(u)$ is the polynomial   (\ref{b18}),
can be consider as the minimum of the function 
\beqa
\label{b21}
G^0_{[x,t, \vc\;]}(u)=xu-3tu^2-F(u)
\eeqa
where $x$, $t$ and   $\vc=(c_0,c_1,\dots,c_{2N+1})$ 
are parameters   and
 $F^{\prime}(u)=f(u)$. 
The minimum is unique only for $t\leq t_c$.
 In \cite{D1} the function of type (\ref{b21}) is extended
to a functional onto the moduli space of all hyperelliptic 
Riemann surfaces
 with real branch points in such a way that the  absolute  
minimum of this functional  gives the solution of the initial 
value problem of the 
Whitham equations (\ref{a2}) for the initial data (\ref{b18}).

First we define the restriction of this  functional
  on the hyperelliptic Riemann surfaces of genus $g$ with
 real  branch points 
 $u_1>u_2>\dots>u_{2g+1}$. The restriction  is a 
function depending on the   branch points. 
This function is build as follows.
Consider the   asymptotic expansion of
 the quasi-momentum $dp_g$ 
(see e.g. \cite{DMN})
\beqa
\label{b22}
dp_g(r)=\left[\dfrac{1}{2\sqrt r}-\dfrac{1}{2\sqrt r} 
\sum_{k=0}^{\infty}
 \dfrac{(2k+1) I_k}{2^{2k+1}r^{k+1}}\right]dr\,.
\eeqa
The coefficients $I_k=I_k(u_1,u_2,\dots,u_{2g+1})$, 
are the so called KdV integrals and are smooth functions of 
the parameter $u_1>u_2>\dots>u_{2g+1}$.
\begin{theo}
\cite{D1} On  the Riemann surface $\cs_g$ consider the function 
\beqa
\label{b23}
G^g_{[x,t, \vc\,]}(\vu) = -x I_0(\vu) + 3 t I_1(\vu) -
\sum_{k=0}^{2N+1}\dfrac{k!}{2^k(2k-1)!!} c_k I_k(\vu)\,,
\eeqa
depending on the variable $u_1,u_2,\dots,u_{2g+1}$ and on the
parameters $x$, $t$ and $\vc$. 
\noindent
Then the equations 
\beqa
\dfrac{\partial}{\partial u_i}
 G^g_{[x,t, \vc\,]}(\vu)=0\;,\quad i=1,2\dots,2g+1\,,
\eeqa
are equivalent to the equations
$\Om_g(r)\mid_{r=u_i}=0\;,\quad i=1,\dots,2g+1,$
where $\Om_g(r)$   has   been defined in (\ref{omega}). 
\end{theo}
\begin{rem}
The function $G^g_{[x,t, \vc\,]}(\vu)$ can also be written 
in the form 
\begin{equation}
\label{gmg}
G^g_{[x,t, \vc\,]}(\vu)=\resi
[dp_g(z)(\cf(z)-2\sqrt z x)], \quad \cf(z)=\int_{0}^{z}
\dfrac{6t\xi+f(\xi)}{\sqrt{z-\xi}}d\xi,
\end{equation}
and $\resi$ is the residue evaluate at infinity.
\end{rem}
\noindent
To extend the function (\ref{b23}) defined on the 
 hyperelliptic
 surfaces 
$\cs_{g\geq 0}$ to a functional on the infinite
 dimensional 
space $M$ of all hyperelliptic Riemann surfaces $\cs_{g\geq 0}$ 
and their degeneration, we follow \cite{D1}.

\noindent
Construct the space $M$  inductively starting from 
$M_0 = \Rn$.
We denote $u$ the coordinate in $M_0$. 
Define now
\beqn 
M_g =M_g^0\cup L_{g-1}\cup T_{g-1}\,,  
\eeqn
where
\beqn
M_g^0 = \left\{ (u_1,u_2, \dots, u_{2g+1}) \in \Rn^{2g+1} |\;
u_1 > u_2 > \dots > u_{2g+1}\right\}\,,
\eeqn
\[
L_{g-1}=\cup_{j=1}^g L_{g-1}^j,\quad T_{g-1}=\cup_{j=1}^g T_{g-1}^j.
\]
Each space $L_{g-1}^j $ is 
 attached to the component of the boundary of $M_g^0$ where
\beqn
u_{2j-1} - u_{2j} \to 0\,,\quad j=1,2,\dots,g\,;
\eeqn
each space $T_{g-1}^j$ is attached to the component of the boundary
of $M_g^0$ where
\beqn
u_{2j}-u_{2j+1}  \to 0 \,,\quad j=1,2\dots,g\,.
\eeqn

\noindent
For reasons that will become clear later, 
we call {\it leading edge} each boundary component $L_{g-1}^j $, 
$j=1,\dots,g$  and {\it trailing edge} 
each  boundary component
$T_{g-1}^j$  $j=1,\dots,g$.

A smooth functional $f$ on $M$ is a sequence of function
$f_g$ defined on every $M_g$ and having   certain good analytic
properties on the boundaries of the space $M_g$ 
(for details cfr \cite{D1}). 
These properties are illustrated in the following example.
%


\noindent
{\bf Example. }
Let us consider the Riemann surface of genus $g+1$
\[
\tilde{\mu}^2=(r-v-\sqrt\d)(r-v+\sqrt\d)
\prod_{j=1}^{2g+1}(r-u_j),\quad v\in\Rn\,,\;\;
u_1>u_2>\dots>u_{2g+1},
\]
where $ v\neq u_j,\;\;j=1,\dots 2g+1$ and $0<\d\ll 1$.
We denote  with  $dp_{g+1}(r,\d)$, the quasi momentum
 defined on this surface and with 
$G^{g+1}_{[x,t, \vc\,]}(\vu,v,\d)$
the function equivalent to  (\ref{b23}) defined on 
this surface and depending on the 
variables $v-\sqrt\d,v+\sqrt\d,u_1,\dots,u_{2g+1}$.  
We study the behavior of  $dp_{g+1}(r,\d)$
 in the limit $\d\rightarrow 0$.

\noindent
First we consider a trailing edge of the boundary of the space
$M_{g+1}$,  namely the case in which 
$v\in (u_{2j},u_{2j-1})$, 
$1\leq j\leq g+1$, $u_{2g+2}=-\infty$.
 We have that  \cite{Fay}
\beqa
\label{dptrail}
dp_{g+1}(r,\d)&=&dp_g(r)+\dfrac{\d}{2}dp_g(v)
\dfrac{\partial}{\partial_v}\om_v(r)+O(\d^2)
\eeqa
where $dp_g(v)=\dfrac{dp_g(r)}{dr}|_{r=v}$ and $O(\d^2)/\d^2 $ 
is a meromorphic differential with zero residues. 
The differential $\om_v(r)$ is a
normalized  Abelian differential of the third kind with 
poles at the points
$P^{\pm}(v)=(v,\pm\mu(v))$ with residue $\pm 1$ respectively, namely
\begin{equation}
\label{third}
\om_v(r)=\dfrac{dr}{\mu(r)}\dfrac{\mu(v)}{r-v}-
\sum_{k=1}^g\dfrac{r^{g-k}dr}{\mu(r)}B_k(v),
\quad \mu^2(r)=\prod_{i=1}^{2g+1}(r-u_i)\,.
\end{equation}
The constants $B_k(v)$ are uniquely determined
 by the normalization conditions
\[
\int_{u_{2k+1}}^{u_{2k}}\om_v(r)=0, \quad k=1,\dots,g.
\]
When 
 $v\in(u_{2j+1},u_{2j})$, $j=0,\dots,g$, and $u_0=+\infty$,  we are 
considering one of the leading edge of the space $M_{g+1}$.

In this case the expansion of the differential
 $dp_{g+1}(r,\d)$ for $\d\ra 0$  reads \cite{Fay}
\begin{equation}
\label{dplead}
dp_{g+1}(r,\d)= dp_{g}(r)+\rho\om_v(r)
\int_{P^-(v)}^{P^+(v)}dp_g(\xi)+O(\rho^2)
\end{equation}
where $\om_v(r)$ has been 
defined in (\ref{third}) and 
$\rho=-\frac{1}{\log\d}$. 
The correction in the right hand side of
(\ref{dptrail}) contains also exponentially small terms like 
$\exp(-1/\rho)$. 

\noindent
Combining equations 
  (\ref{b20}), (\ref{gmg}) and (\ref{dptrail}) we obtain 
the expressions of
the function $G^{g+1}_{[x,t, \vc\,]}(\vu,v,\d)$
 near the trailing  edges  of the space 
$M_{g+1}$, namely 
\begin{equation}  
\label{ft}
G^{g+1}_{[x,t, \vc\,]}(\vu,v,\d)=
G^g_{[x,t, \vc\,]}(\vu)
+\dfrac{\delta}{2}dp_g(v)\Om_g(v)+O(\d^2),
\end{equation}
where the differential $\Om_g(r)$ has been defined in (\ref{omega}) 
and here and below $\Om_g(v)=\frac{\Om_g(r)}{dr}|_{r=v}$.

\noindent
The behavior of
the function $G^{g+1}_{[x,t, \vc\,]}(\vu,v,\d)$
 near the leading  edges  of the space 
$M_{g+1}$,
is obtained from  (\ref{b20}), (\ref{gmg}) and 
 (\ref{dplead}):
\begin{equation}  
\label{fl}
G^{g+1}_{[x,t, \vc\,]}(\vu,v,\d)\simeq G^g_{[x,t, \vc\,]}(\vu)+
\dfrac{ \rho}{2}\int_{P^-(v)}^{P^+(v)}dp_g(r)
\int_{P^-(v)}^{P^+(v)}\Om_g(r).
\end{equation}

 \noindent
The function $ G^g_{[x,t, \vc\,,]}(u_1,u_2,\dots,u_{2g+1})$ 
defined on $M_g^0$
 can be extended to a smooth
 functional on the  space $M$. In \cite{D1} the
extension is built  proving that the 
$I_k(u_1,\dots,u_{2g+1}),\,$ $k\ge 0$,  
can be extended to  smooth functionals on  $M$.
We state the following theorem
\begin{theo}\cite{D1}
The  functional
\beqa
\label{fff}
G_{[x,t, \vc\,]} = -x I_0 + 3 t I_1 -
\sum_{k=0}^{2N+1}\dfrac{k!}{2^k(2k-1)!!} c_k I_k 
\eeqa
is a $C^{\infty}$ smooth  functional on $M$. Its absolute minimum 
 is a $C^1$-smooth
 multi-valued  function of $x$ 
depending $C^1$-smoothly  on the parameters
$t,c_1,\dots,c_{2N+1}$. 

\noindent
If the absolute minimum $(u_1(x,t),\dots,u_{2g+1}(x,t))$ 
belongs to $M_g^0$ for
certain  values of the parameters, then it satisfies the g-phase
Whitham equations. 
\end{theo}


\section{ Study of the function $g=g(x,t)$}
\setcounter{equation}{0}
\setcounter{figure}{0}
\vskip 5pt
In this section we study the behaviour of the genus $g=g(x,t)$
 of the solution of the Whitham equations (\ref{whithamg}) 
for the one-parameter family of  initial data (\ref{bi}).
In \cite{G} it was shown that for polynomial initial data
of degree $2N+1$ the solution of the Whitham equations has
 genus at most equal to $N$.
Hence in the case of  initial data (\ref{bi}) the solution will
have genus at most equal to $2$.
For this reason we need only to describe the locus of  points
 of the $x-t$  plane 
where the genus $g(x,t)$ increases from zero to one or two and 
from one to two. To this end we study the behaviour of 
the functional  (\ref{fff}) on the  phase transition boundaries 
of the space $M_0$, $M_1$ and $M_2$.
The restriction of the functional (\ref{fff}) on $M_0$ has the form
(\ref{b21}). Thus the  solution of the Burgers 
equation (i.e. the Whitham equation for $g=0$)
\begin{equation}
\label{zz}
x=6tu+f(u)
\end{equation}
minimizes the functional (\ref{fff}) for all $x$ and $t<t_c$ where
$t_c$ is the time of gradient catastrophe. 
 Let us  find for which $x$ and $t$ the solution of (\ref{zz})
 is also minimal along directions transversal to $M_0\subset M$.
For the purpose we consider the embedding of $M_0$ as the component of 
the boundaries   of $M_1$ ($T_0$, $L_0$)  and we study the 
behaviour of the functional
(\ref{fff}) near the trailing edge and leading 
edge of the space $M_1$.

\noindent
First we need the following theorem.
\begin{theo} \cite{G2}
On the solution $u_1(x,t)>u_2(x,t)>\dots>u_{2g+1}$ of the $g$-phase
Whitham equations, the differential $\Om_g(r)$ defined in 
(\ref{omega})
reads
\begin{equation}
\label{Om}
\Om_g(r)=-\mu(r)[\partial_r\Psi_g(r,\vu)+\sum_{i=1}^{2g+1}
\partial_{u_i}\Psi_g(r,\vu)]dr
\end{equation}
where the function $\Psi_g(r,\vu)$ is given by the expression
\begin{equation}
\label{Psi}
\Psi_g(r,\vu)=-\resi \left[\dfrac{\cf(z)dz}{\mu(z)(z-r)}\right],
\quad \cf(z)=\int_{0}^{z}\dfrac{6t\xi+f(\xi)}{\sqrt{z-\xi}}d\xi,
\end{equation}
and $\resi$ is the residue evaluate at infinity.
\end{theo}

Using (\ref{ft}), (\ref{Om}) and (\ref{Psi})  
we can simplify the
functional (\ref{fff}) near the trailing edge of 
the space $M_1$ to
the form
\beqa
\label{ktrail}
G^1_{[x,t, \vc\,]}(u,v,\d)= G^0_{[x,t, \vc\,]}(u)+
\dfrac{\delta}{2}(-6t-\partial_v q(u,v)-
\partial_u q(u,v))+O(\d^2)\,,
\eeqa
where $v<u$, $G^0_{[x,t, \vc\,]}(u)$ has been defined in (\ref{b21})
 and
\begin{equation}
q(u,v)=-\resi
\left[\dfrac{\left(\int_{0}^{z}\frac{f(\xi)}
{\sqrt{z-\xi}}d\xi\right)dz}
{(z-v)\sqrt{z-u}}\right]=\dfrac{1}{4\sqrt 2}
\int_{-1}^1\dfrac{f(\frac{1+m}{2}v+\frac{1-m}{2}u)}{\sqrt{1-m}}dm.
\end{equation}
Analogously on the leading edge of the space $M_1$  we obtain
\beqa
\label{klead}
G^1_{[x,t, \vc\,]}(u,v,\d)= G^0_{[x,t, \vc\,]}(u)+16\rho
(v-u)^2(-2t-\partial_u q(u,v))+O(\rho^2)\,.
\eeqa

\noindent
If, for fixed $(u,t)$,  the $\delta$-correction of (\ref{ktrail}) 
is positive for every $v\leq u$, then the minimizer 
belongs to $M_0$. If it is
negative for some values of $v\leq u $, then the 
 minimizer belongs to
the inner part of $M_1$. The points  $v<u$
belong to the trailing edge of the space $M_1$
 if the triple  $(t,u,v)$ is a zero
and a  minimum with respect to $v$ of the $\delta$ correction of
(\ref{ktrail}). From these considerations  we  obtain 
the following lemma.
\begin{lem}
\label{lemtrailing}
The points $v$ and $u$,  $v<u$, belong to the boundary $T_0$ of the
space $M_1$  if
$(t,u,v)$ satisfy the system 
\beqa
\label{trailing}
\left\{\begin{array}{lll}
6t+\partial_v q(u,v)+\partial_u q(u,v)=0&&\\
\partial_v(\partial_v q(u,v)+\partial_u q(u,v))=0&&\\
(\partial_{v})^2(\partial_v q(u,v)+\partial_u q(u,v))<0\,.&&
\end{array}\right.
\eeqa
\end{lem}
The curve $x=x^-(t)$ on the $x-t$ plane where the genus 
increases from zero to one is determined solving the 
above system together with
the zero-phase solution $x=6tu+f(u)$.   We will call the
curve $x^-(t)$ trailing edge.

Analogous consideration can be done for the leading edge.
\begin{lem}
\label{lemleading}
The points $v$ and $u$,  $v>u$, belong to the boundary $L_0$ iff
$(t,u,v)$ satisfy the system
\beqa
\label{leading}
\left\{\begin{array}{lll}
-2t+\partial_u q=0&&\\
\partial_{v}\partial_uq=0&&\\
(\partial_{v})^2 \partial_u q<0\,.&&
\end{array}\right.
\eeqa
\end{lem}
The curve $x=x^+(t)$ on the $x-t$ plane where the genus 
increases from zero to one is determined solving the 
above system together with
the zero-phase solution $x=6tu+f(u)$.    We will call the
curve $x^+(t)$ leading edge.

\begin{rem} Both  systems (\ref{trailing}) and (\ref{leading})  
in the limit $v\rightarrow u$  become
\beqa
\label{zero}
\left\{\begin{array}{lll}
6t+f^{\prime}(u)=0&&\\
f^{\prime\prime}(u)=0&&\\
f^{\prime\prime\prime}(u)<0&&
\end{array}\right.
\eeqa
where $f(u)$ is the initial data (\ref{b18}).
If the above system admits a real solution  $(u,t)$  such that 
\beqa
\label{double}
\left\{\begin{array}{lll}
2t+\partial_u q\le 0\quad \forall\; v\neq u &&\\
 6t+\partial_v q+\partial_u q\le 0\;\quad \forall\; v\neq u\,,&& 
\end{array}\right.
\eeqa
  then $u$ belongs to the boundary $T_0\cap L_0$
of the space $M_1$. The corresponding point
 $(x,t,u)$ is a point of gradient catastrophe of the 
solution of the Burgers
equations.
\end{rem}

\subsection{Boundary between the one-phase solution and the two-phase solution}
In the following, we write explicitly the equations
 determining the phase transition boundary between the one-phase
 solution and the two phase solution. 

\noindent
Let  be $dp_1(r)$ the quasi-momentum  restricted  
to the inner part of $M_1$,
\beqa
\label{dp1}
dp_1(r)=\dfrac{r+\alpha_0}{
\sqrt{(r-u_1)(r-u_2)(r-u_3)}}dr\,,\;\;
 \alpha_0=-\dfrac{\int_{u_3}^{u_2}\frac{rdr}{\sqrt{(r-u_1)
(r-u_2)(r-u_3)}}}
{\int_{u_3}^{u_2}\frac{dr}{\sqrt{(r-u_1)(r-u_2)(r-u_3)}}}
\eeqa
From  (\ref{gmg}) the restriction on the inner part of 
$M_1$ of the functional
 $G_{[x,t, \vc\,]}$  reads
\beqa
\label{gm1}
G^1_{[x,t, \vc\,]}(u_1,u_2,u_3)=\resi
[dp_1(z)(\cf(z)-2\sqrt z x)].
\eeqa
\begin{prop}
The critical points 
 of (\ref{gm1})  on the space of  elliptic curves are 
given  by the equations for $u_1>u_2>u_3$,
\beqa
\label{onep}
x=\lambda_i\,t + w_i\;,\quad i=1,2,3
\eeqa
where $\lb_i=\lb_i(u_1,u_2,u_3) $ has been defined in (\ref{l1}) and
\beqa
\label{sigma}
\begin{array}{lll}
	w_i(u_1,u_2,u_3)=2\dfrac{\prod_{\overset{ j\neq i}{j=1}}^3
	(u_i-u_j)}{\alpha_0+u_i}\partial_{u_i} \sigma(u_1,u_2,u_3)&&\\
\sigma(u_1,u_2,u_3)=-\resi[\dfrac{\left(\int_{0}^{z}\frac{f(\xi)}
{\sqrt{z-\xi}}d\xi\right)dz}
{\sqrt{(z-u_1)(z-u_2)(z-u_3)}}].&&
\end{array}
\eeqa
\end{prop}
\proof
Using the  formula \cite{D1}
\beqa
\dfrac{\partial}{\partial u_i}\alpha_0=-\dfrac{1}{2}+
\dfrac{1}{2}\dfrac{(\alpha_0+u_i)^2}
{\prod_{j\neq i}(u_i-u_j)}\,,\quad i=1,2,3,
\eeqa
equations (\ref{onep}) and (\ref{sigma}) are recovered 
straightforward.
\hfill$\square$

\noindent
The trailing edge of the phase transition boundary  between
 the one-phase
 solution and the two-phase 
solution is determined from (\ref{ft}), (\ref{Om})   and (\ref{Psi}).
When $g=1$ (\ref{Psi}) reads 
\begin{equation}
\Psi_1(r,\vu)=-\resi \left[\dfrac{\cf(z)dz}{(z-r)
\sqrt{(z-u_1)(z-u_2)(z-u_3)}}\right],
\end{equation}
so that (\ref{ft}) simplifies to the form
\begin{equation}  
\label{ft1}
G^{2}_{[x,t, \vc\,]}(\vu,v,\d)\simeq  G^1_{[x,t, \vc\,]}(\vu)+
\dfrac{\d}{2}(v+\alpha_0)(-\partial_v
\Psi_1(v,\vu)-\sum_{i=1}^3\partial_{u_i}\Psi_1(v,\vu))
\end{equation} 
where $G^1_{[x,t, \vc\,]}(\vu)$ has been defined in
(\ref{gm1}) and $\alpha_0$ has been defined in (\ref{dp1}).
 The point
$v\in(-\infty,u_3)$ or $v\in(u_2,u_1)$. In this case 
 $v+\alpha_0\neq 0$.
From the $\d$ correction of (\ref{ft1}) we obtain the 
equations 
determining the trailing edge, namely
\beqa
\label{tone}
\left\{\begin{array}{lll}
\partial_v\Psi_1(v,\vu)+\sum_{i=1}^3
\partial_{u_i}\Psi_1(v,\vu))=0&&\\
&&\\
\partial_v(\partial_v\Psi_1(v,\vu)+\sum_{i=1}^3
\partial_{u_i}\Psi_1(v,\vu))=0&&\\
&&\\
(\partial_v)^2[(v+\alpha_0)(\partial_v\Psi_1(v,\vu)+
\sum_{i=1}^3\partial_{u_i}\Psi_1(v,\vu))<0&&
\end{array}\right.
\eeqa
where  $\vu=(u_1(x,t),u_2(x,t),u_3(x,t))$  is determined from 
(\ref{onep}).
 A similar system
can be obtained for the leading edge using (\ref{klead}).

\noindent
In the limit  $v\rightarrow u_l$, $1\leq l\leq 3$, system 
(\ref{tone}) reads
\beqa
\label{catast}
\left\{\begin{array}{lll}
\partial_{u_l}(\partial_{u_1} \sigma+\partial_{u_2} 
\sigma+\partial_{u_3} \sigma)=0&&\\
&&\\
(\partial_{u_l})^2(\partial_{u_1} \sigma+\partial_{u_2}
\sigma+\partial_{u_3} \sigma)=0&&\\
&&\\
(\partial_{u_l})^3[(\alpha_0+u_l)(\partial_{u_1} 
\sigma+\partial_{u_2}
\sigma+\partial_{u_3} \sigma)]<0&&
\end{array}\right.
\eeqa
where $\s=\s(u_1,u_2,u_3)$ has been defined in (\ref{sigma}).
If the above system admit a solution $x^*,t^*$, then
  the one-phase solution  $u_1(x^*,t^*)>u_2(x^*,t^*)>u_3(x^*,t^*)$ 
determined by 
(\ref{onep}) has on the $u_l$ branch a point of gradient 
catastrophe,
namely $\left.\dfrac{\partial}{\partial x} u_l(x,t)\right|_
{\left\{\overset{x=x^*}{t=t^*}\right\}}
\rightarrow \infty$ and
$\left.\dfrac{\partial^2}{\partial x^2} u_l(x,t)\right|_
{\left\{\overset{x=x^*}{t=t^*}\right\}}
\rightarrow \infty$. 
\begin{theo}
\label{ttheo}
If the initial data (\ref{b18}) satisfies the condition 
$f^{\prime\prime\prime}(u)<0$,  
then the solution of the one-phase Whitham
equations  has no point of gradient catastrophe. 
\end{theo}
\proof
Using the following expression for 
$\sigma(u_1,u_2,u_3)$ 
  \cite{FRT1} 
\beqa
\label{q}
\sigma(u_1,u_2,u_3)=\dfrac{1}{2\sqrt 2\pi}\int_{-1}^{1}
\int_{-1}^1 \dfrac{f\left(\frac{1+s}{2}\frac{1+t}{2}u_1+
\frac{1-s}{2}\frac{1+t}{2}u_2+\frac{1-t}{2}u_3\right)}
{\sqrt{(1-t)(1-s^2)}}dt\,ds\;,
\eeqa
it can be easily checked that the second equation of (\ref{catast})
cannot be satisfied for any real  $u_1,u_2$ and $u_3$.
\hfill $\square$

\begin{rem} Using a different approach  Tian has proved 
 \cite{FRT1}that 
for smooth monotone  decreasing
initial data  satisfying the
condition $f^{\prime\prime\prime}(u)<0$, the solution of the 
one phase Whitham equations exists
for all $t>0$.
\end{rem}

\section{ Bifurcation diagrams}
We   study the bifurcation diagram of the solution  of the Whitham
equations for the one parameter family of initial data
\beqa
\label{id}
f_c(u)=-(u^3+c\,u^4+u^5)\;,\quad c^2\leq \dfrac{15}{4}\,.
\eeqa
For such initial data, the functional (\ref{fff}) reads
\beqa
\label{a11}
G_{[x,t, c]}=-xI_0+3t I_1-\dfrac{1}{20}I_3-\dfrac{1}{70}\ c I_4-
\dfrac{1}{252}I_5\;.
\eeqa
The restriction of this functional  on $M_0$ (that is on the 
curve $\mu^2=r-u$, $\;u\in\Rn$) has the form
\beqa
\label{a12}
G^0_{[x,t, c]}(u)=xu-3tu^2+ \dfrac{u^4}{4}+ c \dfrac{u^5}{5}+
\dfrac{u^6}{6}\;,
\eeqa
thus the minimizer given by
\beqa
\label{a13}
x=6ut- u^3-c u^4-u^5
\eeqa
solves the Burgers equation until the time of gradient 
catastrophe $t_0=0$.
At later times the minimizer of $G_{[x,t,c]}$ may belong 
to $M_0$, $M_1$ or $M_2$.

\noindent
The  Burgers equation  has another point of gradient catastrophe if 
 (\ref{zero}) and (\ref{double})  with the initial data 
(\ref{id}) are satisfied,  that is
\beqa
\left\{\begin{array}{lll}
\label{czero}
-6 t +3u^2+4 cu^3+5 u^4=0&&\\
6 u +12 c u^2 +20 u^3=0&&\\
6+24 c u +60 u^2>0&&\\
6t+\partial_v q_c+\partial_u q_c\le0 \quad\forall\; v\neq u\,,
\quad v\in\Rn\,,&&\\
2t+\partial_u q_c\le0 \quad\forall\; v\neq u\,,\quad v\in\Rn\,,&&
\end{array}\right.
\end{eqnarray}
 where
\begin{equation}
\label{qc}
\begin{split}
q_c(u,v)=& -1/35\,(5\,u^3 + 6\,u^2\,v + 8\,u\,v^2 + 16\,v^3)
     - c/315\,( 35\,u^4 + \\
&40\,u^3\,v + 64\,u\,v^3+
        48\,u^2\,v^2 + 128\,v^4) 
 - 1/693\,(63\,u^5 + \\
      &  70\,u^4\,v +80\,u^3\,v^2 + 96\,u^2\,v^3 + 128\,u\,v^4 + 
      256\,v^5)\,.
\end{split}
\end{equation} 
The solutions of (\ref{czero}) are obviously $u_0=0,\;t_0=0$ and
\beqa
\left\{\begin{array}{lll}
\label{a31}
u_1= -\frac{3}{10}(c +\sqrt{ c^2-\frac{10}{3}})\;\;\; \mbox{for} 
\;c>0\;,&&\\
u_1= -\frac{3}{10}(c -\sqrt{ c^2-\frac{10}{3}})\;\;\; \mbox{for} 
\;c<0\;,&&\\
t_1=\frac{1}{6}(3  u_1^2+4 c u_1^3+ 5 u_1^4)>0\,,&&\\
\end{array}\right. 
\end{eqnarray}
with the constraints 
\beqa
\label{intw}
-\dfrac{\sqrt{15}}{2}\leq
c\leq-\dfrac{1}{6}\sqrt{5(13+5\sqrt{7})}\,,\;\;\dfrac{1}{2}
\sqrt{\frac{1}{11}(75+21\sqrt{15})}\leq
c\leq
\dfrac{\sqrt{15}}{2}\,,
\eeqa
which are  obtained from the last two inequalities of (\ref{czero}).
We put
\beqa
\label{nu1} 
&&\nu_1=-\dfrac{1}{6}\sqrt{5(13+5\sqrt{7})}\simeq -1.90863\\ 
&&\nu_4=\sqrt{\frac{1}{44}(75+21\sqrt{15})}\simeq 1.88494
\label{nu4}
\eeqa
 because these numbers occur frequently in the following. 
The $x_1$ coordinate
 relative to $(t_1,u_1)$ 
 is recovered from the equation (\ref{a13}). For $c$ positive 
$x_1>0$ and for
 $c$ negative $x_1<0$. 

\noindent Here and below all the numerical results are obtained
using Mathematica 3.0 for Solaris
Copyright 1988-97 Wolfram Research, Inc.

\subsection{Trailing edges\label{T}}
The equations determining the trailing edge of the space $M_1$
are given by 
  system (\ref{trailing}), namely
\beqa
\left\{\begin{array}{lll}
\label{a19}
6t +\partial_v q_c+\partial_u q_c=0&&\\
\partial_v(\partial_v q_c+\partial_u q_c)=0&&\\
(\partial_{v})^2(\partial_v q_c+\partial_u q_c)<0\;,&&
\end{array}\right.
\eeqa
where $q_c(u,v)$ has been defined in (\ref{qc}).
The above system can be easily solved numerically 
getting $u=u(t)> v=v(t)$.
Substituting the value of $u=u(t)$ in the zero phase solution
(\ref{a13}) we get, in the $x-t$ plane,  the curve $x^-(t)$ which
 describes the phase transition from the $g=0$ solution to 
the $g=1$ solution. For some values of $c$  and $t$ 
system (\ref{a19})
has two different solutions $u_1(t)>v_1(t)$ and $u_2(t)>v_2(t)$ and
correspondingly we have two curves $x_1^-(t)$ and $x_2^-(t)$
in the $x-t$ plane. This occurs for $c$ in the intervals
\[
-\dfrac{\sqrt{15}}{2}\leq c<\nu_1,\quad \nu_4< c\leq  
\dfrac{\sqrt{15}}{2}
\]
where $\nu_1$ and $\nu_4$ have been  defined in (\ref{nu1}) and
(\ref{nu4}) respectively.

\begin{figure}[htb]
\hskip 1.5cm
\centering\psfig{figure=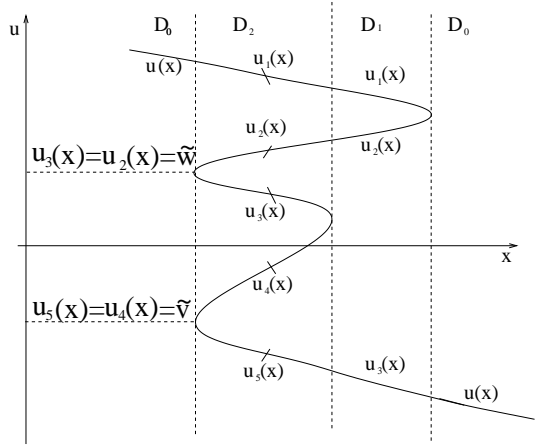,width=7cm}
\caption{Double trailing  edge.}
\label{tesitrail}
\end{figure}

A   particular solution
of system (\ref{a19})  is the one which describes the
double-trailing edge (cfr. below).
 System (\ref{a19}) determines  
the zeros which are also 
a  maxima with respect to $v$ of the polynomial 
$6t +\partial_v q_c+\partial_u q_c$.   
Since $\partial_v q_c+\partial_u q_c$ is a fourth degree
 polynomial in the
variable  $v$  with negative leading
coefficients, it cannot  have  more than two maxima.  
When system (\ref{a19}) admits such two maxima 
$(\tilde{t},\tilde{u},\tilde{v})$ 
 and $(\tilde{t},\tilde{u},\tilde{w})$ 
with $\tilde{t}>0$, $\tilde{v}\leq \tilde{u}$ and 
$\tilde{w}\leq \tilde{u}$,  we are in 
a situation of a double
trailing edge (see
Figure~\ref{tesitrail}). The 
degenerate  Riemann surface  describing this situation
is given by the equation 
$\mu^2=(r-\tilde{u})(r-\tilde{v})^2(r-\tilde{w})^2$ and 
represents the
boundary $T_1^1\cap T_1^2$ of the space $M_2$.
The curve $x^-(t)$ describing the trailing edge in the
$x-t$ plane looses the $C^1$-smoothness at the point $t=\tilde{t}$.
\noindent
Such two solutions
exists  for $c$ in the intervals  
\beqn
-\frac{\sqrt{15}}{2}\leq c \leq\nu_1\;, \quad
\nu_3\leq
c\leq \frac{\sqrt{15}}{2} \,,
\eeqn
where $\nu_1$ has been defined in (\ref{nu1}) and 
\beqa
\label{nu3} 
\nu_3=\sqrt{ \frac{5}{2} + \frac{5}{18}\left(
\frac{25}{18}\right)^{\frac{1}{3}}
\left((27-7\sqrt{21})^{\frac{1}{3}}+(27+7\sqrt{21})^{
\frac{1}{3}}\right)}\simeq 1.78167\,.
\eeqa

\noindent
For  $c=\nu_3$ the two  solutions
$(\tilde{t},\tilde{u},\tilde{v})$ 
 and $(\tilde{t},\tilde{u},\tilde{w})$ 
   become coincident, namely  $\tilde{v}=\tilde{w}$.  

\noindent
For  $c=\nu_1$ 
the solution $(\tilde{t},\tilde{u},\tilde{w})$  satisfies  
 $\tilde{w}=\tilde{u}$. 
The point $(\tilde{v},\tilde{w}=\tilde{u})$ belongs  to 
 the component $L_1^1\cap T_1^1\cap T_1^2  $ of the boundary
of $M_2$ and the corresponding degenerate Riemann surface is 
$\mu^2=(r-\tilde{u})^3(r-\tilde{v})^2$. 
The solution of the Burgers equation has in
 $(\tilde{x},\tilde{t},\tilde{w}=\tilde{u})$ 
a point of gradient catastrophe in correspondence of
 the trailing edge.


\noindent
For $c$ in the interval
$\nu_1 < c <\nu_3$
there exists just one real solution $v(t,c)< u(t,c)$ of system
(\ref{a19})  for all $t>0$. Substituting $u(c,t)$ in the
zero-phase equation we get a  curve $x^-(t)$ in the $x-t$ plane 
which is smooth for all $t>0$.

\subsection{Leading edges\label{L}}

The leading edges of the space $M_1$ 
 are determined from system (\ref{leading})  namely
\beqa
\left\{\begin{array}{lll}
\label{b57}
2t +\partial_u q_c(u,v)=0&&\\
\partial_v\partial_u q_c(u,v)=0&&\\
(\partial_{v})^2\partial_u q_c(u,v)<0&&\
\end{array}\right.
\eeqa 
where $q_c(u,v)$ has been defined in (\ref{qc}).
For some values of  $t>0$ and for $c$ in the intervals 
\[
-\dfrac{\sqrt{15}}{2}\leq c<\nu_1,\quad \nu_4< c\leq  
\dfrac{\sqrt{15}}{2}
\]
system (\ref{b57})
has two different 
solutions $u_1(t)<v_1(t)$ and $u_2(t)<v_2(t)$.
Correspondingly we have two leading edges $x_1^+(t)$ and $x_2^+(t)$
in the $x-t$ plane which describe the phase transition from
the $g=0$ solution to the $g=1$ solution. 

\begin{figure}[htb]
\begin{minipage}[h]{.43\linewidth}
\centering\epsfig{figure=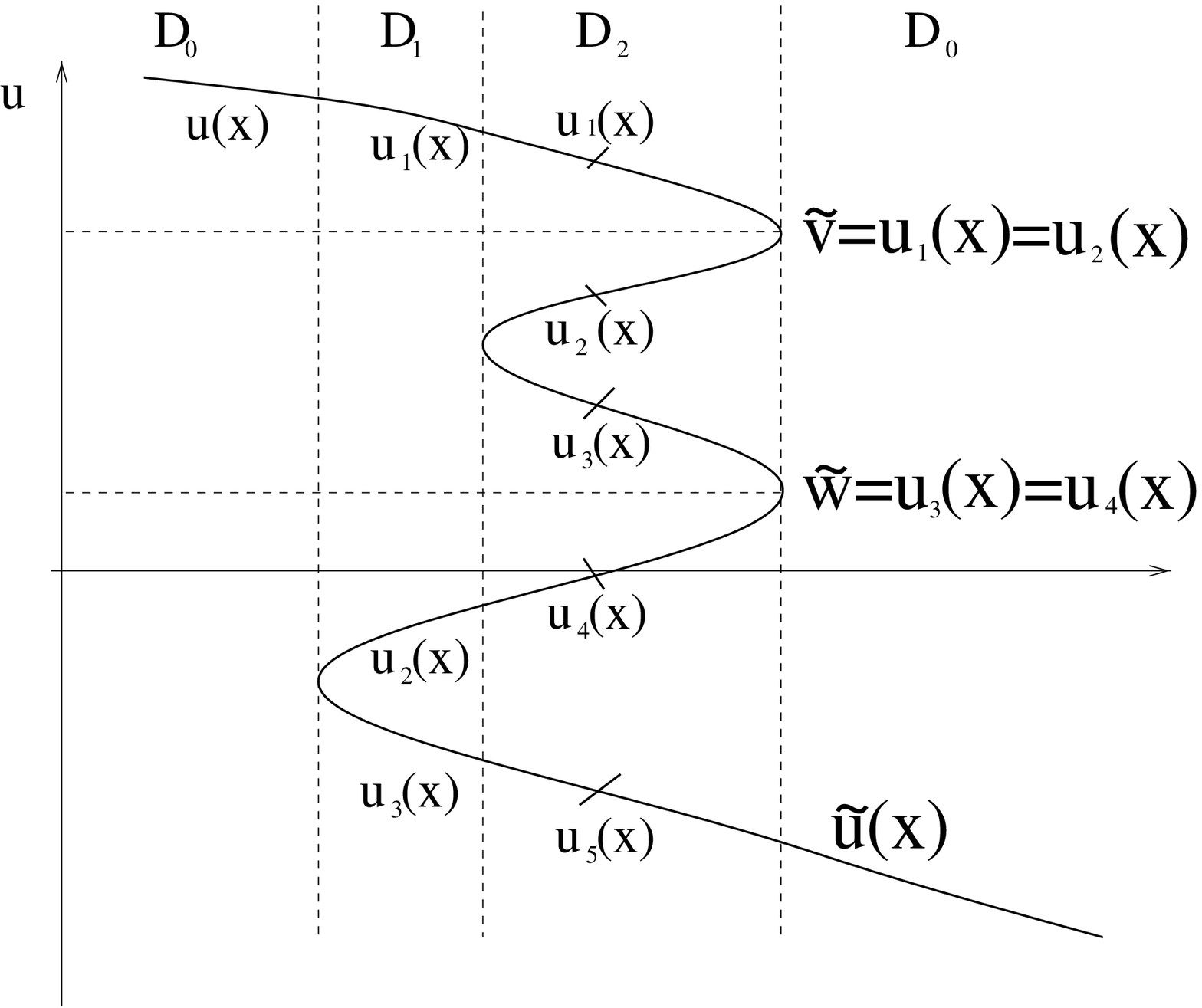,width=\linewidth}
\vskip .2cm
\caption{Double leading edge.\label{c222}}
\end{minipage} \hskip 1.0cm
\begin{minipage}[h]{.43\linewidth}
\centering\hskip 30pt\epsfig{figure=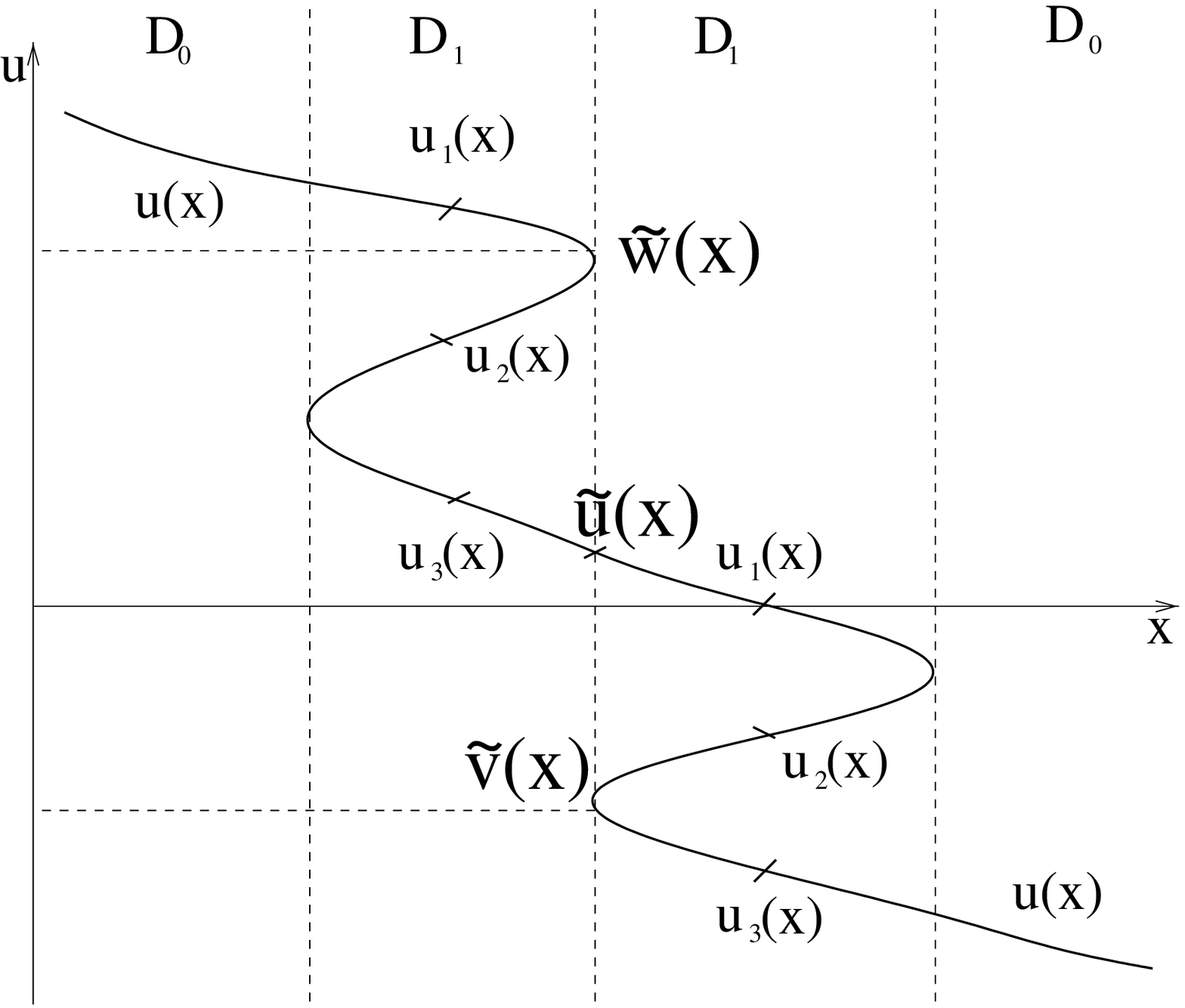,width=\linewidth}
\caption{Leading-Trailing edge.\label{figlt}}
\end{minipage}
\end{figure}

\noindent
We also  consider the situation of a double leading 
edge as shown in Fig.\ref{c222}. That is we 
 study  for which values of   $c$ system (\ref{b57}) has
 two real  solutions $(\tilde{t},\tilde{u},\tilde{w})$ and 
$(\tilde{t},\tilde{u},\tilde{v})$ 
  satisfying the constraints $\tilde{v}\geq
\tilde{u},\;\;\tilde{w}\geq 
\tilde{u}$ and $\tilde{t}>0$.
 These solutions
 belong to the boundary component $L_1^1\cap L_1^2$ 
 of the space $M_2$ which 
is described  by the degenerate Riemann surface 
$\mu^2=(r-\tilde{u})(r-\tilde{v})^2(r-\tilde{w})^2$.
This double solution occurs only for $c$ in the intervals
\beqn
\nu_4\leq c\leq\sqrt{\frac{15}{4}}\,,\quad
-\sqrt{\frac{15}{4}}\leq c<\nu_2\,,
\eeqn
where $\nu_4$ has been defined in (\ref{nu4}) and  
\vskip 2pt
\beqa
\label{nu2}
\nu_2=-\sqrt{
\dfrac{5}{2} +\dfrac{35}{11}\cos{\frac{\pi-\theta}{3}}}
\simeq -1.85585,\;\;\cos\theta=\dfrac{11}{14}\,.
\eeqa

\noindent
For $c=\nu_4$, these two solutions become coincident, namely 
 $\tilde{v}=\tilde{w}$. 
 
For $c=\nu_2$ the solution $(\tilde{t},\tilde{u},\tilde{w})$ 
satisfies the  relation  $\tilde{w}=\tilde{u}$. 
The  corresponding  point 
$(\tilde{x},\tilde{t},\tilde{w}=\tilde{u})$ is a point of gradient
catastrophe of the Burgers equation and it occurs at 
the leading edge.  

\noindent
When $c\in (\nu_2,\;\nu_4)$, 
there exists just one real solution $v(t,c)>u(t,c)$ of 
system (\ref{a19})
 for all $t>0$. In this case there exists just a single leading
 edge for all $t>0$. 
\subsection{Leading-trailing edge\label{LT}}
We call leading-trailing edge the situation in which  a leading edge and a
trailing edge have the same $(x,t)$ coordinates as shown in Fig.\ref{figlt}.
\noindent
In other words a leading trailing edge is determined by a solution
$(\tilde{t},\tilde{u},\tilde{w})$ of system (\ref{b57}) and 
$(\tilde{t},\tilde{u},\tilde{v})$ of system (\ref{a19}) where 
$\tilde{w}\geq\tilde{v}$.
We  check numerically that such two solutions exist for   $c$ in the intervals
\[
-\sqrt{\frac{15}{4}}\le  c < \nu_1\,,\quad
\nu_4< c\leq \sqrt{\frac{15}{4}}\,.
\]

\subsection{Point of gradient catastrophe of  the one-phase solution}

The solution of the one-phase Whitham 
equation with the initial data (\ref{id}) has a point of gradient catastrophe
if the correspondent system (\ref{catast})  has a solution.
From Theorem~\ref{ttheo}  it follows that  for $c^2<\frac{5}{2}$, the solution
of the one-phase Whitham equations with initial data (\ref{id}) 
has  no point of gradient catastrophe.

We solve numerically  systems (\ref{catast}) and (\ref{onep})
   for the initial data (\ref{id})
restricting $c$ in the intervals  $-\sqrt{15}/2\leq c\leq -\sqrt{5/2}$
 and  $\sqrt{5/2}\leq c\leq
\sqrt{15}/2$.

\begin{figure}[hb]
\centering
\mbox{\subfigure[]{\psfig{figure=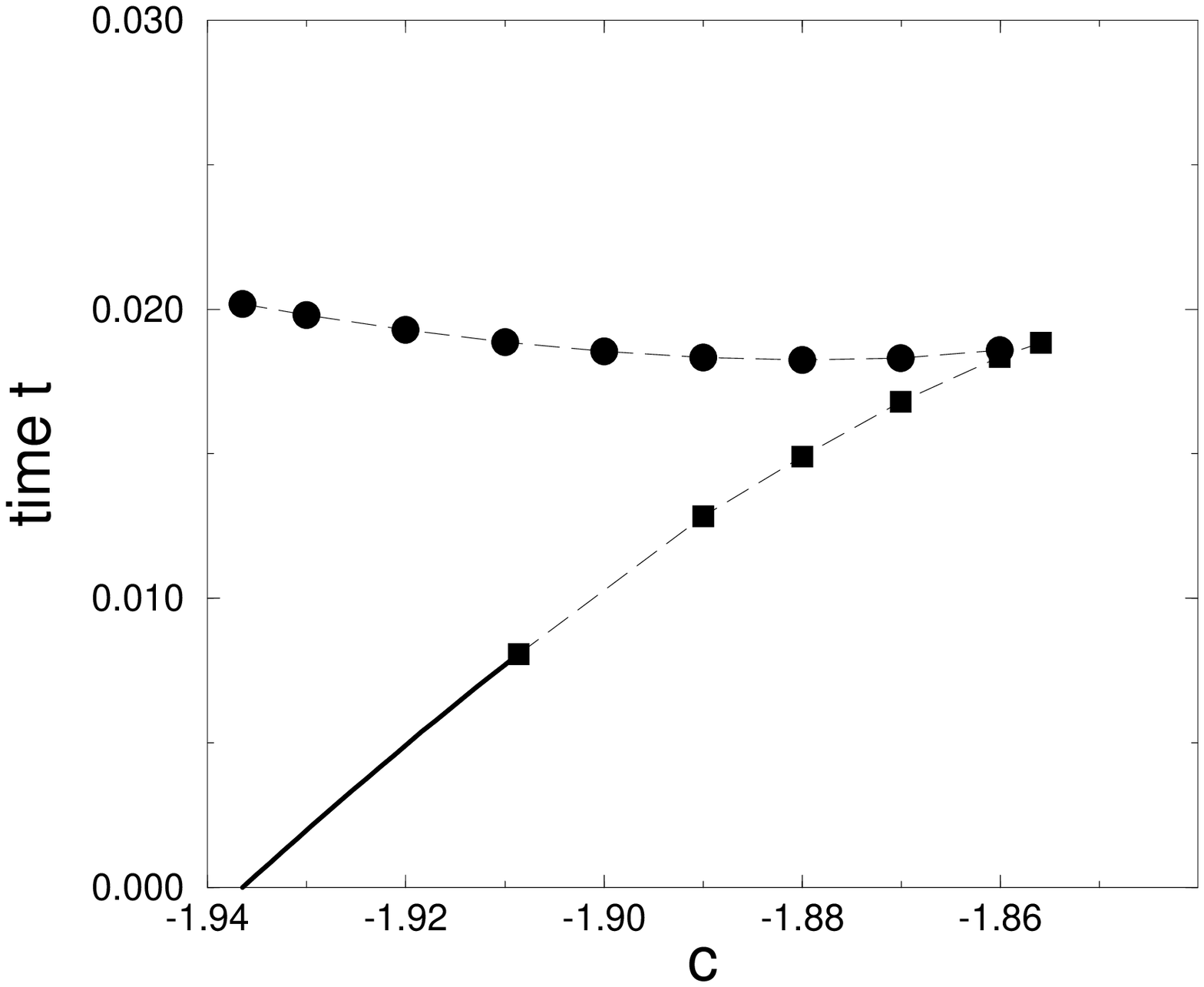,width=.5\textwidth}}\quad
\subfigure[]{\psfig{figure=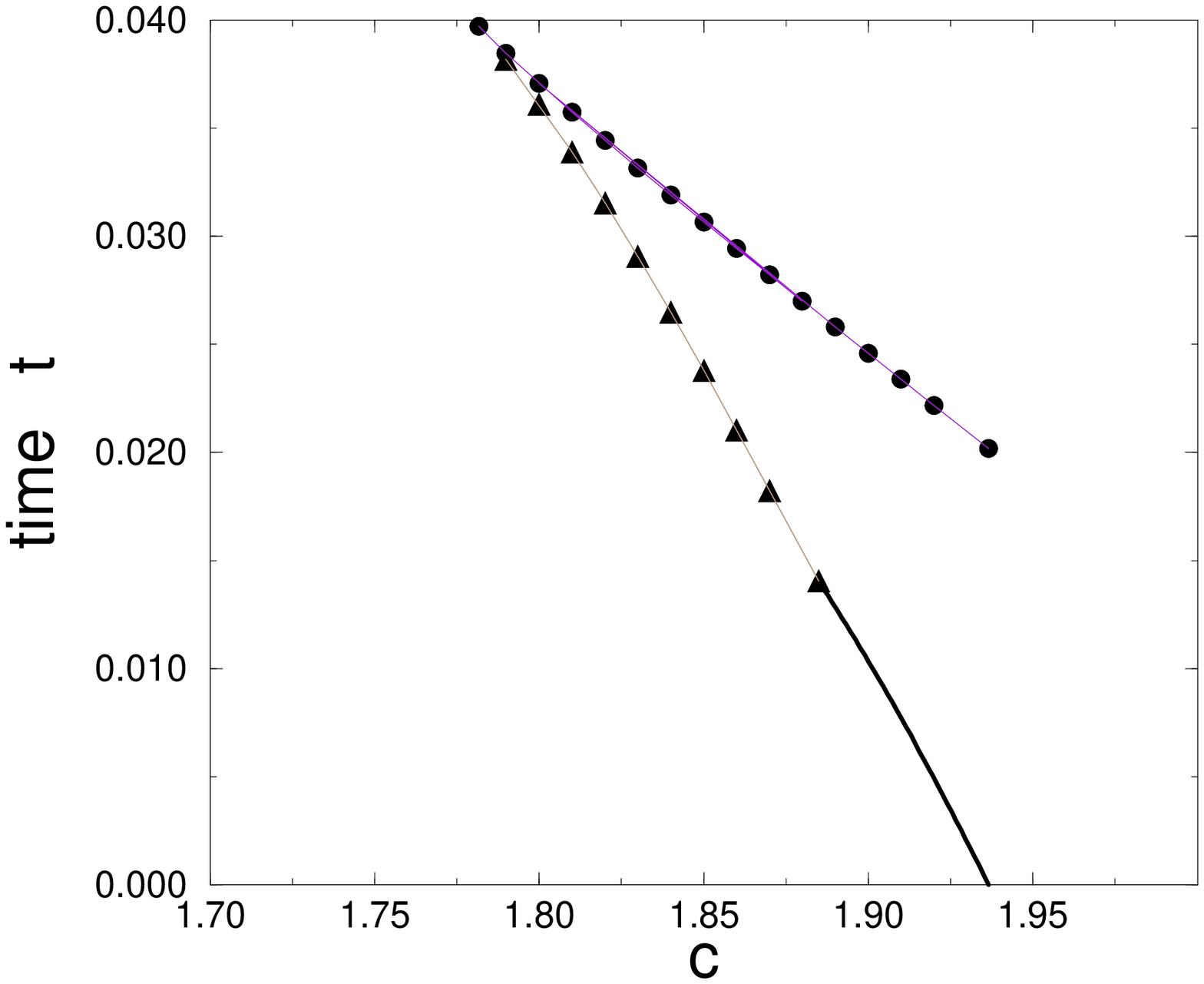,width=.5\textwidth}}}\quad
\caption{On the plots a) and  b) the time of gradient catastrophe
of the one-phase solution 
is plotted as a function of $c$. The circles represent the points of gradient
catastrophe on the $u_2$ branch, the triangles are the points of gradient
catastrophe on the $u_3$ branch and the squares are the points of gradient
catastrophe on the $u_1$ branch. The solid  lines represent the time
of gradient catastrophe of the zero-phase solution as  a function of $c$.} 
\label{figc2}
\end{figure}

We find that there exists a point of
gradient catastrophe for the solution of the one-phase Whitham 
equations on  the $u_1$-branch  for $c\in[\nu_1, \nu_2)$,
where $\nu_1\simeq -1.90863$ and $\nu_2\simeq -1.85585$ have been
defined in (\ref{nu1}) and (\ref{nu2}) respectively, 
(see figure~\ref{figc2}).

There is a point of gradient catastrophe in the solution of the
one-phase Whitham equations on the $u_2$-branch for $c$ in the
intervals $[-\frac{\sqrt{15}}{2},\nu_2)$ and
$(\nu_3,\frac{\sqrt{15}}{2})$,
see figure~\ref{figc2}.

There is a point of gradient catastrophe in the solution of the
one phase Whitham equations on the $u_3$-branch for
$c\in(\nu_3, \nu_4]$,
where $\nu_3\simeq 1.78167$ and $\nu_4\simeq 1.88494$ have
 been defined in (\ref{nu1}) and (\ref{nu4})
respectively, see figure~\ref{figc2}. 

\begin{rem}
A point of gradient catastrophe appears in the solution of the one-phase
Whitham equations whenever the solution is changing genus.
If the point of gradient catastrophe appears on the $u_1$ or $u_3$-branch,
the corresponding solution will increase genus by one near this point. \\
If the point of  gradient catastrophe appears in the one-phase solution 
on the $u_2$-branch it means that the
two-phase solution has just disappeared.
We have checked numerically this fact solving  system (\ref{tone}) in a
neighborhood of each point of gradient catastrophe. 
System (\ref{tone})  determines the phase transition boundary 
between the one-phase
solution and the two-phase solution. 
When a point of gradient catastrophe $(x^*, t^*)$ appears on the 
$u_1$ or $u_3$-branch, then system (\ref{tone}) admits a solution for
$t>t^*$. Namely a two-phase oscillatory zone is developing.
On the contrary if   a point of gradient catastrophe $(x^*, t^*)$ 
appears on the $u_2$ branch, then system (\ref{tone}) does not have
 admissible   solutions for
$t>t^*$. 
\end{rem}

\subsection{Bifurcation diagrams in the $x-t$ plane}
\setcounter{equation}{0}
We  draw in the $(x,t)$ plane the various topological types
of bifurcation diagrams  of the solution of the Whitham
equations with initial data
$x=-u^3-c\,u^4-u^5\,,\;$  $c^2\leq \frac{15}{4}$.
We draw with a solid line the points of the $(x,t)$ plane where the 
genus increases from zero to one  and with a dashed line 
 the points where the genus increases from one to two. We have several cases.

\vskip 10pt
\noindent
1) $\nu_4< c\leq \sqrt{\frac{15}{4}}$;
there exists  a second breakpoint for the zero-phase solution
 in $x_1>0$, $t_1>0$; there exists 
 a point of gradient catastrophe on the $u_2$-branch of the one-phase solution for $t>t_1$;
 there are  a double trailing edge,
a leading-trailing edge and a double leading edge 
hence the bifurcation diagram of
the genus $g(x,t)$ is
\vskip 15pt
\hskip 1.5cm
\hbox{
\vbox{\hsize 150pt
\centering{\psfig{figure=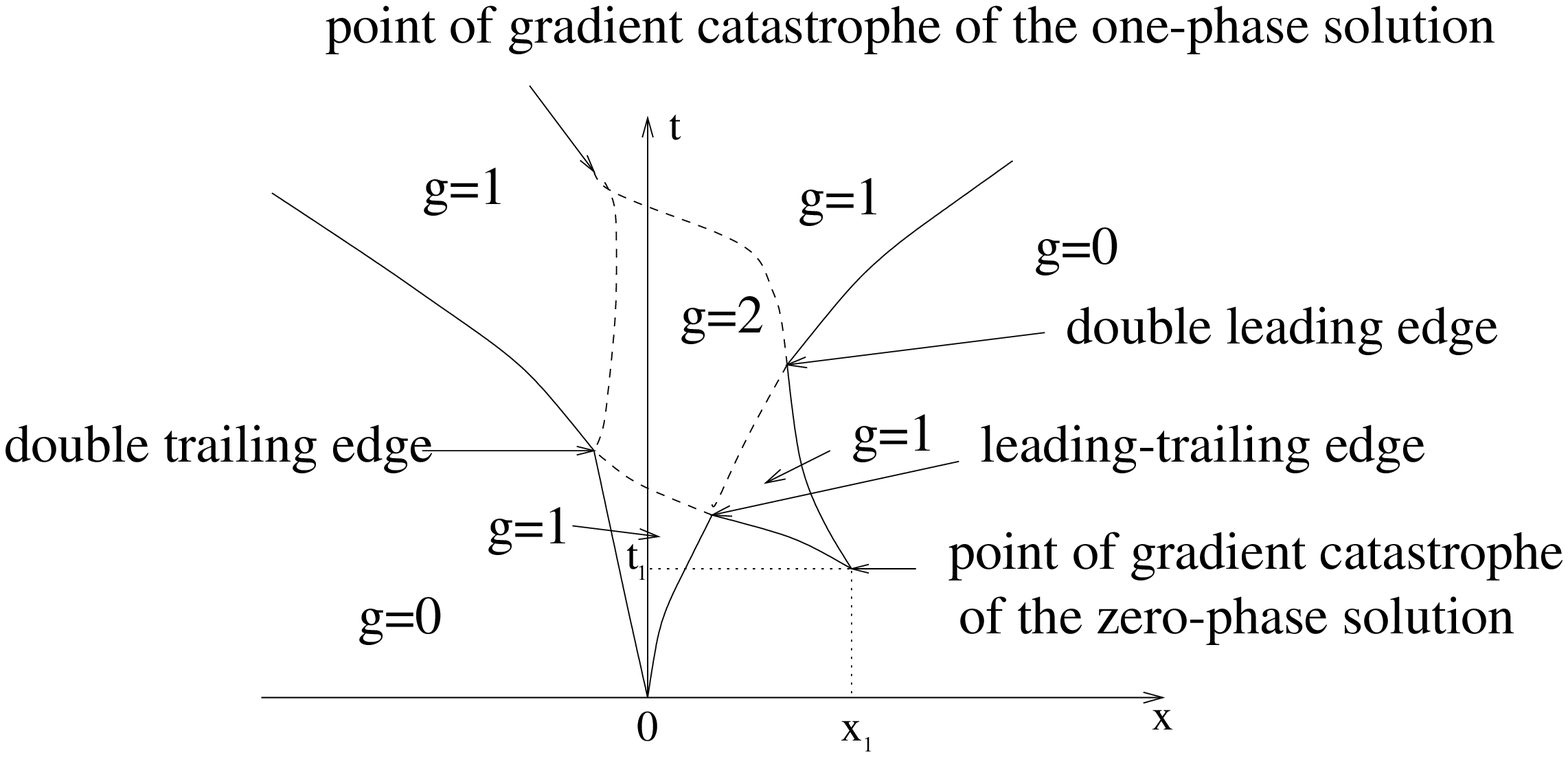,height=2.0in }}

}
}
\vskip 0.7cm

\noindent
2a) $\,c=\nu_4$; the Burgers equation has a second point of gradient catastrophe
in correspondence of the leading edge, there is a point of gradient catastrophe on the $u_2$-branch of the one-phase solution  and there
 is a double trailing edge.

\noindent
2b) $\,c=\nu_1$; the solution of the Burgers equation has a 
second point of gradient catastrophe  $ (x_1,t_1)$
in correspondence of the trailing edge, there  exists a point of gradient
catastrophe on the $u_2$-branch of the one-phase solution for $t>t_1$
and there is a double leading edge.
\vskip 0.5cm
\begin{minipage}[h]{.5\linewidth}
\centering\psfig{figure=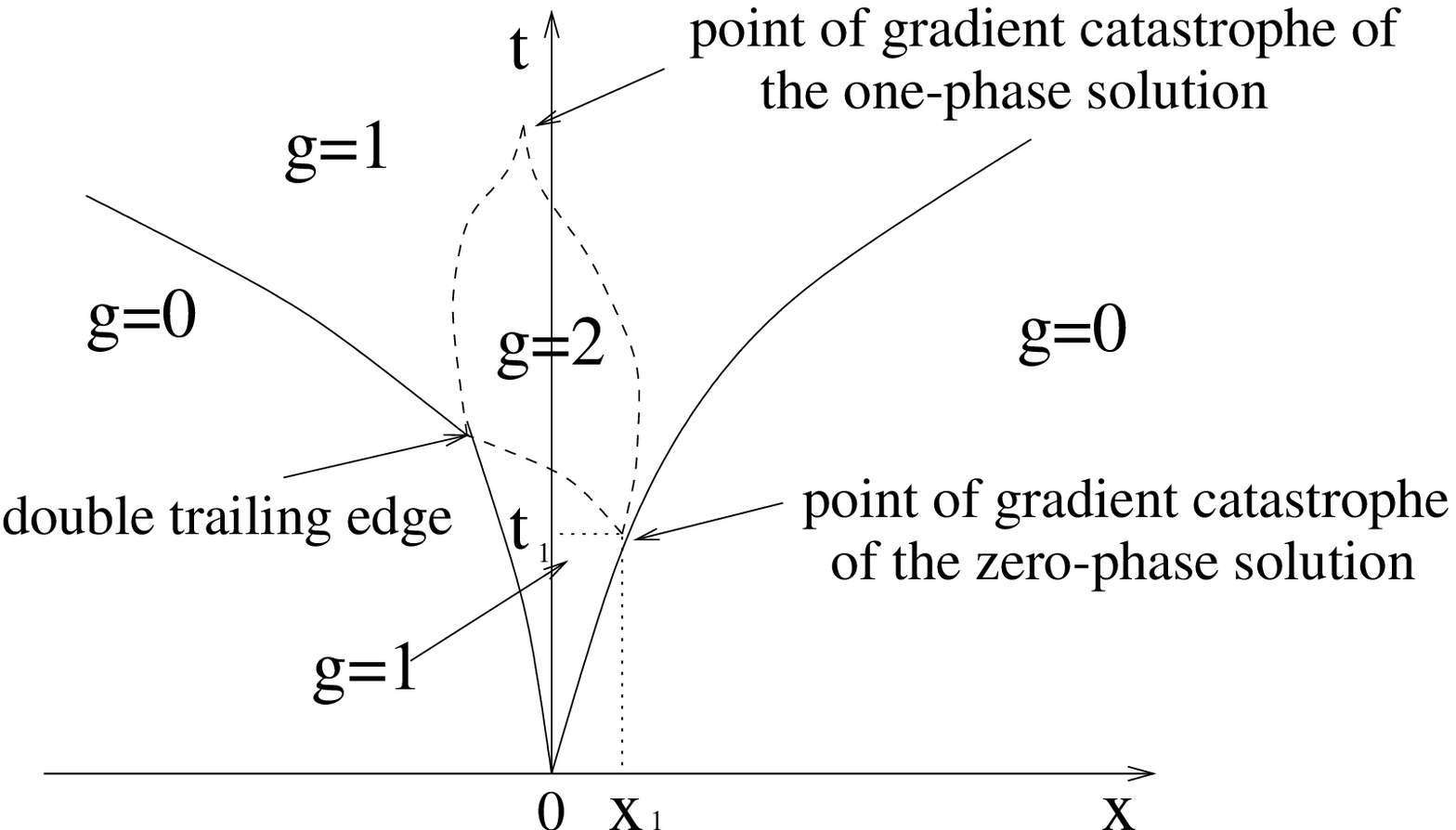,height=2.0in }
\centerline{2a)   }
\end{minipage}
\begin{minipage}[h]{.5\linewidth}
\centering\psfig{figure=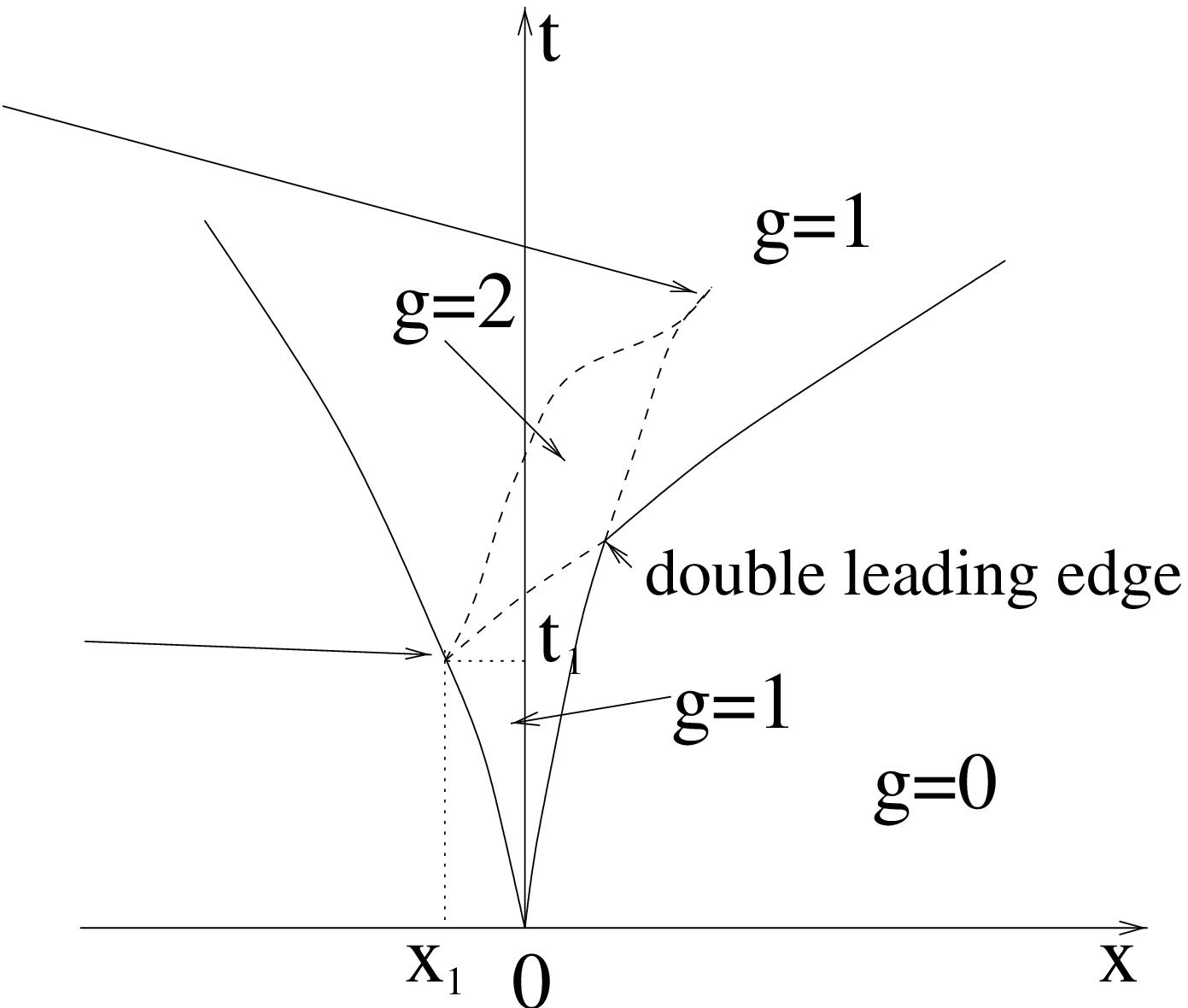,height=2in }
\centerline{ 2 b) }
\end{minipage}
\vskip 1cm

\noindent
 3a) $\,\nu_1<c < \nu_2$; there are  points of gradient catastrophe in the one-phase solution on the $u_1$-branch and $u_2$-branch 
 for $t>0$  and there is double leading edge.

\noindent
3b)
$\nu_3<c< \nu_4\,;$ there are  points of gradient catastrophe in the
the one-phase solution  on the $u_3$-branch and one on the $u_2$-branch 
  and there is a double trailing edge.
\vskip 0.5cm

\begin{minipage}[h]{.5\linewidth}
\centering\psfig{figure=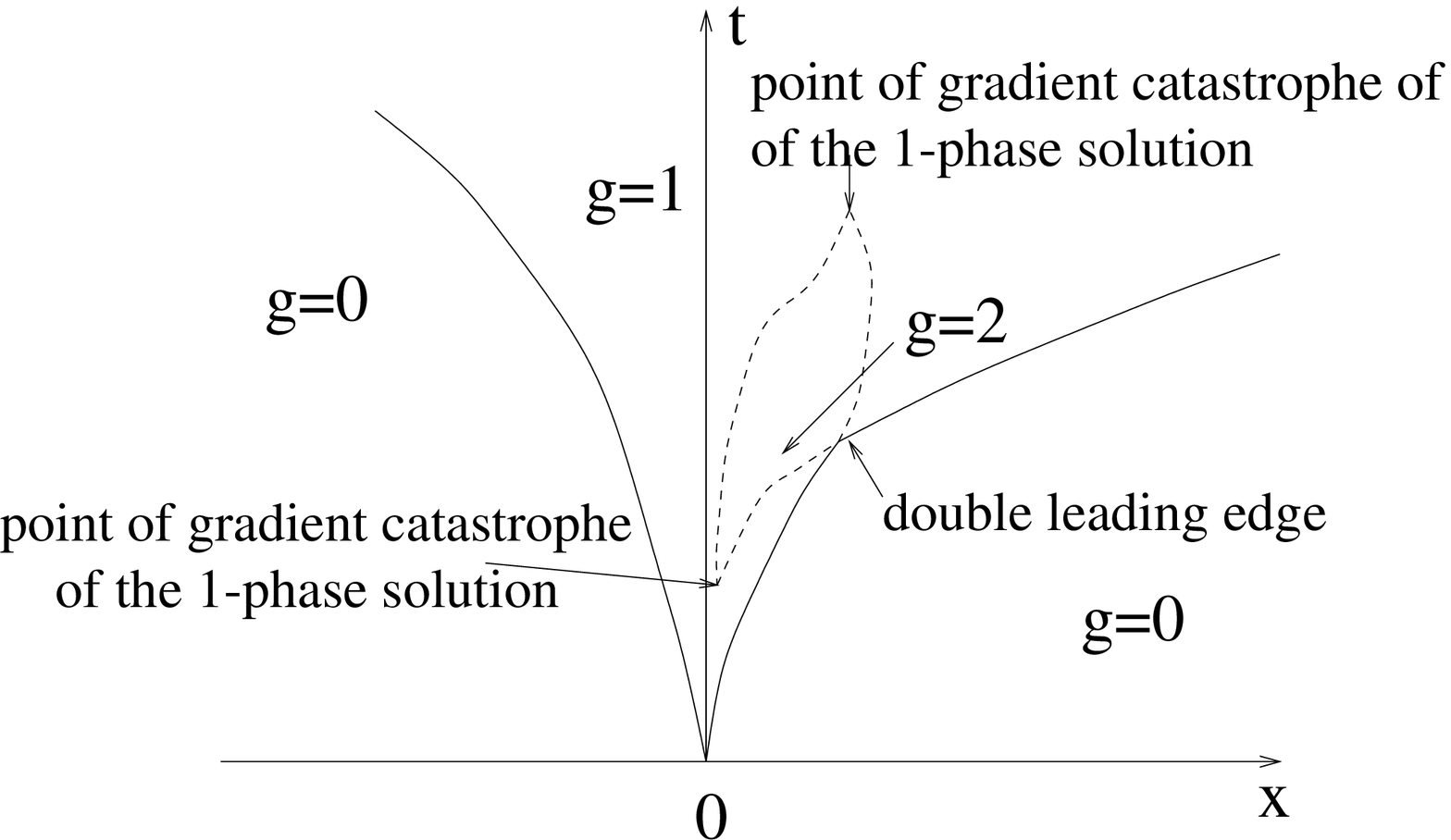,height=2in }
\centerline{3a)   }
\end{minipage}
\begin{minipage}[h]{.5\linewidth}
\centering\psfig{figure=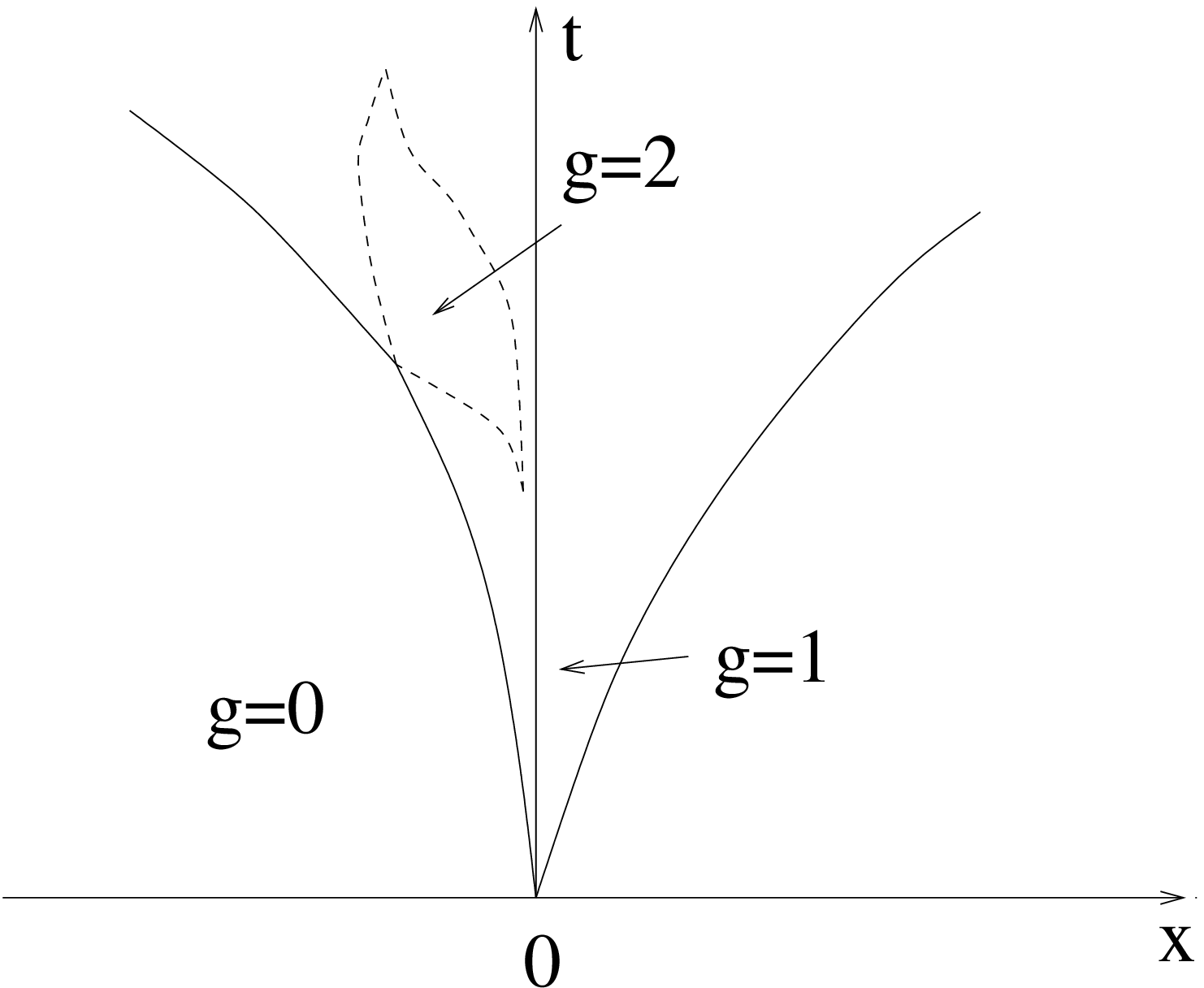,height=2in }
\centerline{  3b) }
\end{minipage}
\vskip 0.7cm

\noindent
4) $\nu_2\leq c \leq\nu_3$; there is only the point $(x=0,t=0,u=0)$
 of gradient catastrophe of the solution
of the Burgers equation. The curves $x^-(t)$ and $x^+(t)$ are $C^1$-smooth
for all $t>0$.
\vskip 3pt
\hbox{
\hskip 60pt
\vbox{\hsize 140pt
\centering{\psfig{figure=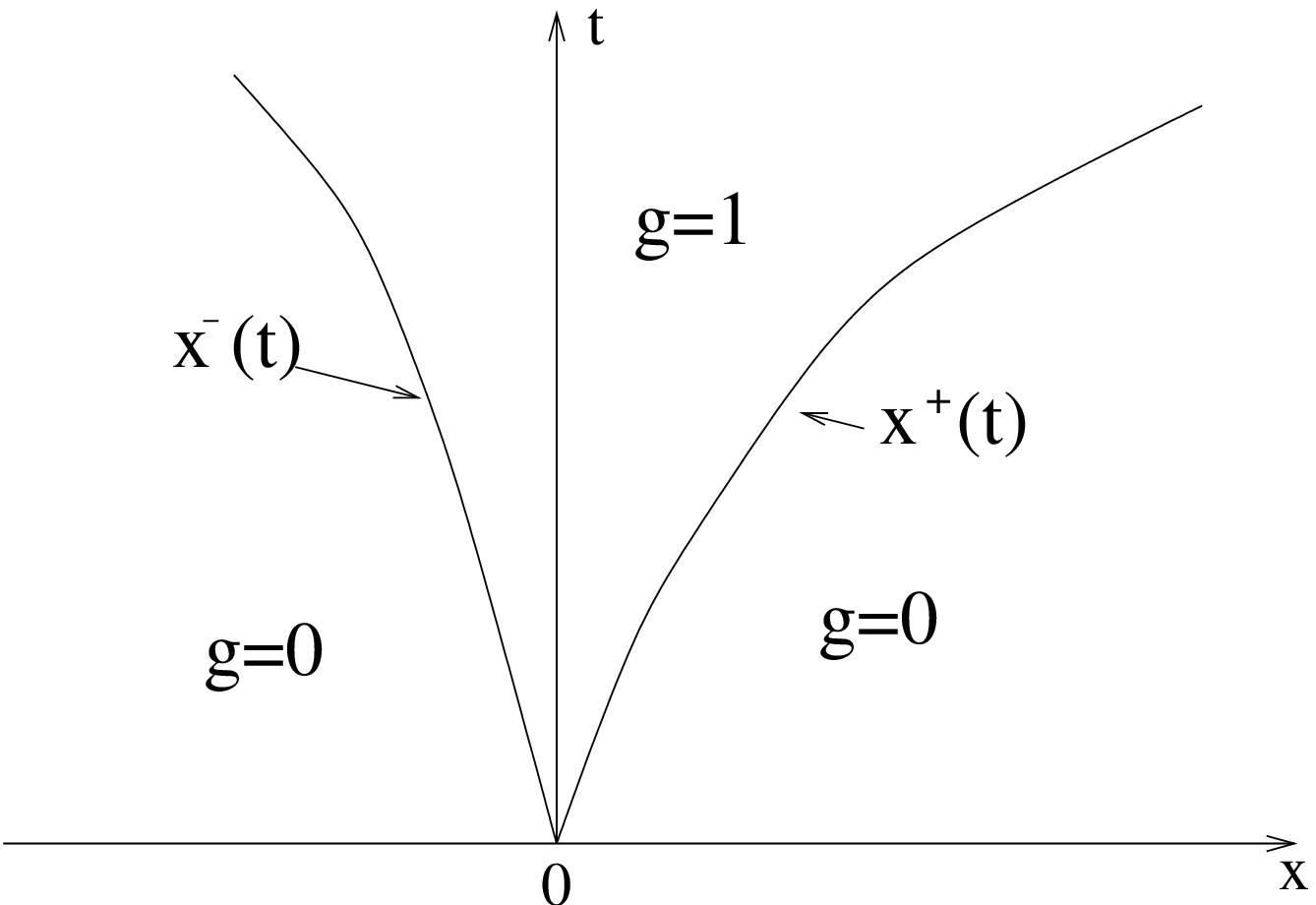,height=1.7in }}
}
}
\noindent
For $c=\nu_3$ the curve
 $x^-(t)$ looses the $C^1$-smoothness for a single value of $t> 0$.
  For $c=\nu_2$ the  curve $x^+(t)$ looses the $C^1$-smoothness for a single value of $t> 0$.
\vskip 0.4cm
\noindent
5) $-\sqrt{\frac{15}{4}}\leq c<\nu_1$; there is a second breakpoint in the
 solution of the 
Burgers equation for $x_1<0$ and $t_1>0$; there is a point of gradient catastrophe in the one-phase solution on the $u_2$-branch for $t>t_1$;  there are a double leading edge,
a leading-trailing edge and a double trailing edge.

\vskip 7pt
\hbox{
\hskip 60pt
\vbox{\hsize 140pt
\centering{\psfig{figure=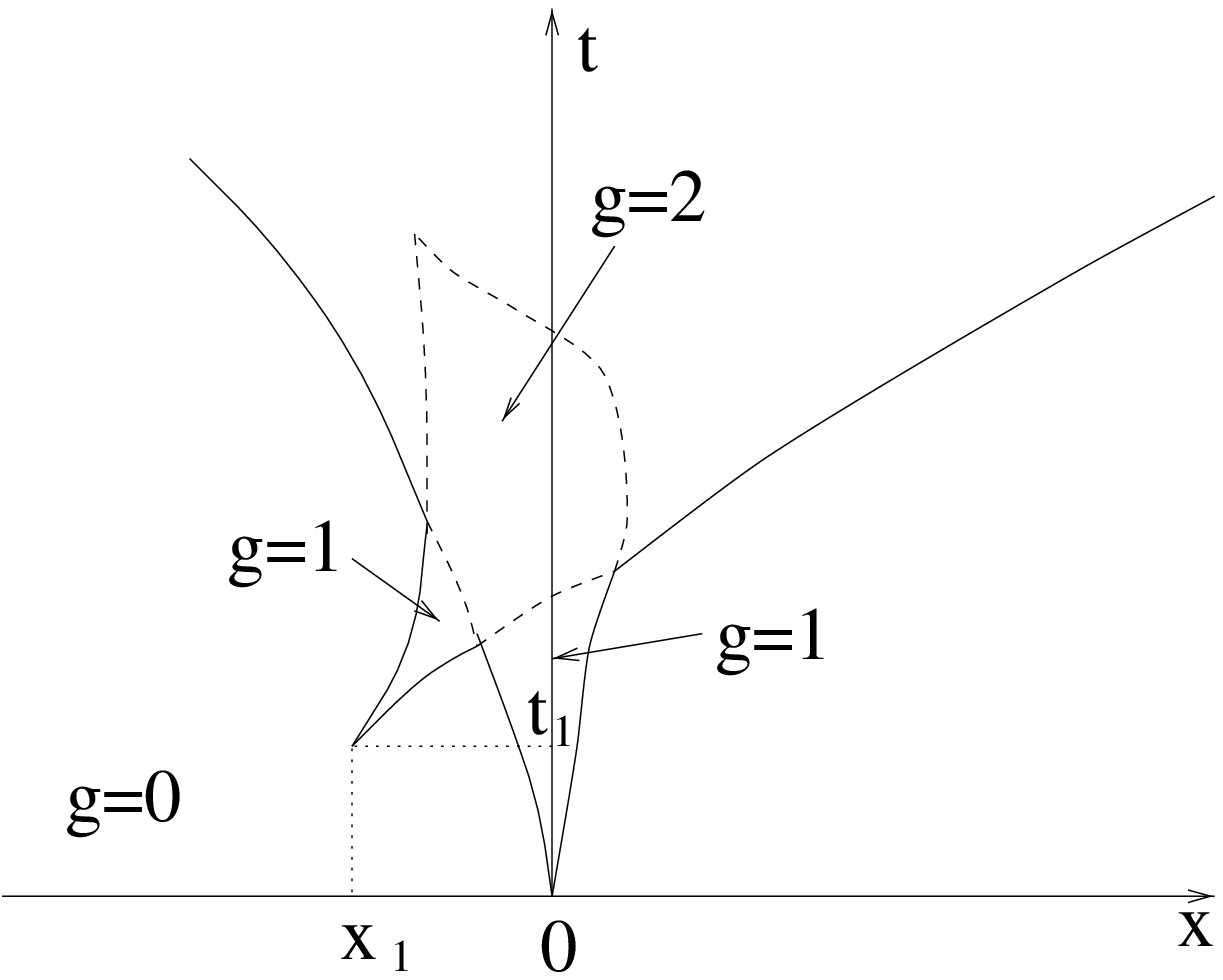,height=2.0in }}
}
}
\subsection{Conclusion}
In this paper we have studied the bifurcation diagram of the solution
of the Whitham equations 
for a one-parameter
family of initial data. On the $x-t$ plane we have characterized the 
bifurcation diagrams in terms
of particular singular points which we have 
called double leading edge,
double trailing edge, leading-trailing edge 
and points of gradient catastrophe. The $x-t$ plane is split in
regions where the solution of the Whitham equations has genus
$g=0,1,2$.
The genus $g=2$ solution survives only  for a finite time.
This result is in agreement with the result in \cite{G2},\cite{FRT3} where
it is  shown that  the solution of the Whitham
equations with monotone polynomial initial data 
has a universal one-phase  self-similar asymptotics. 
This implies that   the large time behaviour   
 of the bifurcation  diagram of
   the solution of the Whitham equations can be described analitically. 
Indeed, as follows from \cite{G2},  
for all times $t>T$, where $T>0$ is a sufficiently
large time, the solution of the Whitham equations for the one
parameter family of initial data (\ref{bi}) is
of genus one inside the interval $x^-(t)<x<x^+(t)$, 
where $x^-(t)\simeq -16.82 t^{\frac{4}{5}}$ and
$x^+(t)\simeq 1.58 t^{\frac{4}{5}} $. It is of 
genus zero outside this interval.

\vskip 1cm

{\bf Acknowledgments.} I am indebted to    Professor
 Boris Dubrovin  who posed
me the problem of this work and gave me many suggestions for its
 solution.

\end{document}